\documentclass{aa} 

\usepackage{graphicx}
\usepackage{txfonts}
\usepackage{natbib}
\bibpunct{(}{)}{;}{a}{}{,}

\newcommand{\be}{\begin{equation}}
\newcommand{\ee}{\end{equation}}
\newcommand{\angstrom}{\mbox{\normalfont\AA}}

\def\apjl{ApJL}
\def\apjs{ApJS}

\newcommand{\Mpc}{$h^{-1}$\thinspace Mpc}

\newcommand{\vmh}{h^{-1}\mathrm{Mpc} }

\begin{document}  

\title{
Supercluster A2142 and collapse in action - infalling and merging 
groups and galaxy transformations  
}

\author {Maret~Einasto\inst{1} 
\and Mirt~Gramann\inst{1} 
\and Changbom~Park\inst{2}
\and Juhan~Kim\inst{3}
\and Boris~Deshev\inst{1,4}
\and Elmo~Tempel\inst{1,5} 
\and Pekka~Hein\"am\"aki\inst{6}
\and Heidi~Lietzen\inst{1}
\and Anne~L\"ahteenm\"aki\inst{7,8} 
\and Jaan~Einasto\inst{1,9,10}
\and Enn~Saar\inst{1,9}
}
\institute{Tartu Observatory, University of Tartu, Observatooriumi 1, 61602 T\~oravere, Estonia
\and
School of Physics, Korea Institute for Advanced Study, 85 Hoegiro, Dong-Dae-Mun-Gu, Seoul 02455, Korea
\and
Center for Advanced Computation, Korea Institute for Advanced Study, 
Hoegiro 87, Dong-dae-mun-gu, Seoul, 130-722, Korea
\and
Institute of Physics, University of Tartu, W. Ostwaldi 1, 50411, Tartu, Estonia
\and
Leibniz-Institut f\"ur Astrophysik Potsdam (AIP), An der Sternwarte 16, D-14482
Potsdam, Germany
\and 
Tuorla Observatory, University of Turku, V\"ais\"al\"antie 20, Piikki\"o, Finland
\and
Aalto University, Mets\"ahovi Radio Observatory, Mets\"ahovintie 114, FI-02540 Kylm\"al\"a, Finland
\and
Aalto University Department of Electronics and Nanoengineering, P.O. Box 15500, FI-00076 Aalto, Finland
\and
Estonian Academy of Sciences, Kohtu 6, 10130 Tallinn, Estonia
\and
ICRANet, Piazza della Repubblica 10, 65122 Pescara, Italy
}

\authorrunning{Einasto, M. et al. }

\offprints{Einasto, M.}

\date{ Received   / Accepted   }

\titlerunning{SCl~A2142}

\abstract
{
Superclusters with collapsing cores represent
dynamically evolving environments for galaxies, galaxy groups, and clusters.
}
{We study the dynamical state and properties of galaxies and groups
in the supercluster SCl~A2142 that has a collapsing core, 
to understand its possible formation and evolution.
}
{We find the substructure of galaxy groups using normal mixture modelling. We 
have used the projected phase space (PPS) diagram, 
spherical collapse model, clustercentric distances, and magnitude gap between
the brightest galaxies in groups to study the dynamical
state of groups and to analyse group and galaxy properties.
We compared the alignments of
groups and their brightest galaxies with the supercluster axis. 
}
{The supercluster core has a radius of about $8$~\Mpc\ and total 
mass $M_{\mathrm{tot}} \approx 2.3\times~10^{15}h^{-1}M_\odot$
and is collapsing. 
Galaxies in groups on the supercluster axis have older stellar populations
than off-axis groups, with median stellar ages $4 - 6$ and $< 4$~Gyr,
correspondingly. 
The cluster A2142 and the group Gr8 both host galaxies  with the oldest stellar populations
among groups in SCl~A2142 having the  median stellar age $t > 8$~Gyr. 
Recently quenched galaxies and active galactic nuclei (AGNs) are mostly located 
at virial radii or in merging regions of groups, and at clustercentric distances 
$D_c \approx 6$~\Mpc.
The most elongated groups lie
along the supercluster axis and are aligned with it.   
Magnitude gaps between the brightest galaxies 
of groups are less than one magnitude, suggesting that
groups in SCl~A2142 are dynamically young.
}
{
The collapsing core of the supercluster, infall of galaxies and groups, and possible 
merging groups, which affect galaxy properties 
and may trigger the activity of AGNs, show how the whole supercluster
is evolving. 
}


\maketitle

\section{Introduction} 
\label{sect:intro} 

The formation of the cosmic web of galaxies, galaxy groups,
clusters, and superclusters connected by galaxy filaments 
started from the tiny 
density perturbations in a very early Universe
\citep{1978MNRAS.185..357J, 1988Natur.334..129K}. 
The studies of protoclusters, progenitors of the present-day
galaxy clusters, have shown that they start to form in regions
of the highest density in the cosmic density field and 
are present already at redshifts $z \approx 6$ 
\citep[][and references therein]{2016ApJ...826..114T, 2016A&ARv..24...14O,
2018MNRAS.474.4612L}. Galaxy protoclusters are the first sights
of galaxy formation in a  high redshift Universe
\citep{2017ApJ...844L..23C, 2018Natur.553...51M, 2018ApJ...861...43P}.
Rich galaxy clusters grow through merging and accretion 
of smaller structures (galaxies and groups of galaxies) along filaments 
\citep[][and references therein]{1996Natur.380..603B, 2009LNP...665..291V,
2011A&A...531A.149S, 2012ARA&A..50..353K, 2014MNRAS.441.2923C}. 
Simulations show that the present-day rich galaxy clusters 
have assembled  half of their mass since redshift $z \approx 0.5$, and
they continue to grow \citep{2013ApJ...763...70W, 2013ApJ...779..127C, 2015JKAS...48..213K,
2015ApJ...806..101H, 2017A&A...607A.131D, 2018A&A...610A..82E}.

In the cosmic web the most luminous galaxy clusters 
are typically located in the high-density core
regions of rich superclusters  \citep{Joeveer:1978a,1978MNRAS.185..357J}.
The high-density cores of rich superclusters 
are the largest objects that may collapse 
now or during future evolution
\citep{1998ApJ...492...45S, 
2000AJ....120..523R, 2006A&A...447..133P,
2014MNRAS.441.1601P, 2015A&A...581A.135G, 2015MNRAS.453..868O, 2015A&A...575L..14C, 
2016A&A...595A..70E}. These cores form an evolving environment
for the study of the formation, evolution and present-day 
properties of galaxies, groups, and clusters inside them.

In this paper we study the dynamical state and
properties of galaxies and galaxy groups in the 
supercluster SCl~A2142 embedding the very rich galaxy cluster \object{A2142}
\citep{2015A&A...580A..69E}. 
The core of this supercluster is already collapsing 
\citep{2015A&A...580A..69E, 2015A&A...581A.135G}.
In the collapsing
cores of superclusters we can study various processes
in groups and clusters  
which are responsible for transformation
of galaxies from mostly blue, star-forming field galaxies to red, quiescent 
cluster population \citep[see][for a review and references]
{2015ApJ...806..101H, 2017ApJ...843..128R}.
\citet{2018A&A...610A..82E} studied the properties of the most 
massive cluster in this supercluster, A2142. In this paper
we extend this analysis to the whole supercluster.
The main goal of this paper is to study the structure and galaxy content
of the supercluster SCl~A2142, 
and in this way to understand its possible formation and evolution.
We have analysed whether  the supercluster environment and infall of galaxies and groups into 
the main cluster of the supercluster, A2142, affects their properties. 
We find substructure of galaxy groups using normal mixture modelling, and 
apply the projected phase space (PPS) diagram, the spherical
collapse model, a magnitude gap between
the brightest galaxies in groups, and clustercentric distances
to study the dynamical
state of galaxy groups in the supercluster  and analyse galaxy properties
at various clustercentric distances. 
We have compared alignments of
groups and their brightest galaxies, and the supercluster axis. We assumed the standard cosmological parameters: the Hubble parameter $H_0=100~ 
h$ km~s$^{-1}$ Mpc$^{-1}$, matter density $\Omega_{\rm m} = 0.27$, and 
dark energy density $\Omega_{\Lambda} = 0.73$ 
\citep{2011ApJS..192...18K}.

\section{Data} 
\label{sect:data} 

\subsection{Supercluster, group, and filament data}
\label{sect:gr}

Our galaxy, group, and supercluster data are based on  
the MAIN sample of the tenth data release of the Sloan Digital Sky 
Survey \citep[SDSS, ][]{2011ApJS..193...29A, 2014ApJS..211...17A}.
We used a spectroscopic galaxy sample  with the 
apparent Galactic extinction corrected $r$ magnitudes $r \leq 
17.77$ and redshifts $0.009 \leq z \leq 0.200$. 
We corrected the redshifts of galaxies for the motion relative 
to the cosmic microwave background and 
computed the comoving distances of galaxies \citep{2002sgd..book.....M}. 
Galaxies with unreliable parameters (large galaxies, which may
have multiple entries in the sample, bright over-saturated
stars that are classified as galaxies, and so on)
were removed from the sample as described in detail 
in \citet{2012A&A...540A.106T, 2014A&A...566A...1T}.

The SDSS MAIN dataset was used to calculate the luminosity-density
field to detect superclusters of galaxies as connected high-density volumes,
to find groups of galaxies with the friends-of-friends algorithm, 
and to determine galaxy filaments by applying the Bisous process to the
distribution of galaxies \citep{2014MNRAS.438.3465T, 2016A&C....16...17T}. 
The data from supercluster, group, and filament catalogues were then
used to select galaxy,  group, and filament  information for the 
supercluster SCl~A2142.
As in \citet{2015A&A...580A..69E}, galaxies are considered as members of
filaments if their distance from the filament axis is up to $0.8$~\Mpc.

The galaxy luminosity density field was calculated using
$B_3$ spline kernel with the smoothing length 8~\Mpc:
\begin{equation}
    B_3(x) = \frac{1}{12} \left(|x-2|^3 - 4|x-1|^3 + 6|x|^3 - 4|x+1|^3 + |x+2|^3\right).
\end{equation}
We created a set 
of density contours by choosing a density threshold and defined connected 
volumes above a certain density threshold as superclusters. In order to choose 
the proper density level for determining individual superclusters, we analysed the 
properties of the density field superclusters at a series of density levels. 
As a result we used 
the density level $D8 = 5.0$
(in units of mean density, $\ell_{\mathrm{mean}}$ = 
1.65$\cdot10^{-2}$ $\frac{10^{10} h^{-2} L_\odot}{(\vmh)^3}$)
to determine individual superclusters.
At this density level superclusters in the richest 
chains of superclusters in the volume under study  still form separate systems. 
At lower density levels they join into huge percolating systems. 
The calculation of the luminosity density field and determination of
superclusters is described in detail in   \citet{2012A&A...539A..80L}. 

\begin{figure}[ht]
\centering
\resizebox{0.47\textwidth}{!}{\includegraphics[angle=0]{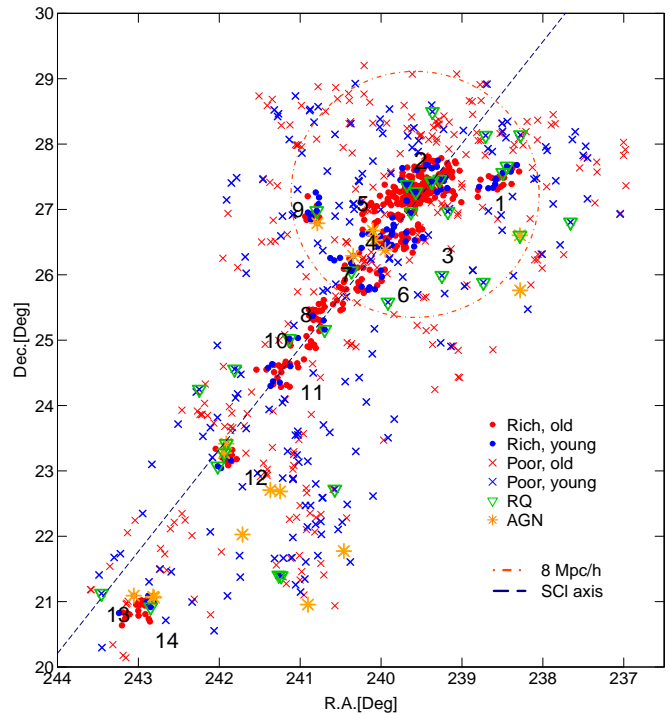}}
\caption{
Sky distribution of galaxies in SCl~A2142. 
The red symbols show galaxies with old stellar populations
($D_n(4000) \geq 1.55$), and the blue symbols denote galaxies with young 
stellar populations ($D_n(4000) < 1.55$). Filled circles denote 
galaxies in  rich groups 
with at least ten member galaxies. The crosses denote galaxies in poor groups 
with less than ten galaxies.
The green triangles show 
recently quenched  galaxies (RQ)
with $D_n(4000) \leq 1.55$ and star formation rate $\log \mathrm{SFR} < -0.5$.
Orange stars indicate the positions of AGNs (see text for definitions).
 The red-orange circle shows a projected radius 
of about $8$~\Mpc.
The dashed line shows the supercluster axis.
}
\label{fig:radec}
\end{figure}

{\it Supercluster catalogue and supercluster SCl~A2142.}
Data on SCl~A2142 are taken from the catalogue
of galaxy superclusters by \citet{2012A&A...539A..80L}, in which
SCl~A2142 at redshift $z \approx 0.09$ has over a thousand member galaxies
within the SDSS MAIN sample. 
The total length of this supercluster, defined as the maximum distance between 
galaxy pairs
in the supercluster, is $\approx 50$~\Mpc\ \citep{2012A&A...539A..80L}.
This supercluster was recently described 
in \citet{2015A&A...580A..69E} and \citet{2015A&A...581A.135G}.
They showed that SCl~2142 has a collapsing high-density core 
in an almost spherical main body with a 
radius of about $11$~\Mpc,  
and an almost straight tail. We used this definiton for the 
main body and tail of the supercluster also in this paper.

{\it Galaxy groups  in the supercluster SCl~A2142.}
We selected galaxy groups in SCl~A2142 
from the group catalogue by \citet{2014A&A...566A...1T}, where
galaxy groups were  determined using 
the friends-of-friends (FoF) cluster analysis 
method introduced in cosmology by \citet{1982Natur.300..407Z} and 
\citet{1982ApJ...257..423H}. A galaxy belongs to a 
group of galaxies if this galaxy has at least one group member galaxy closer 
than a linking length. In a flux-limited sample the density of galaxies slowly 
decreases with distance. To take this selection effect into account properly 
when constructing a group catalogue from a flux-limited sample, the 
linking length was rescaled with distance, calibrating the scaling relation by observed 
groups. As a result, the 
maximum sizes in the sky projection and the velocity dispersions of 
\citet{2014A&A...566A...1T} groups 
are similar at all distances. 
Details of the data 
reduction, group finding procedure, and description of the group catalogue can be found in 
\citet{2014A&A...566A...1T}.

It is possible that our group finding algorithm combines
together close groups, especially in high density regions of supercluster
cores. Also, it may sometimes split or fragment single groups into several
smaller groups \citep[fragmentation of groups, see][]{2015MNRAS.453.3848D}.
Various group finding algorithms and mass estimation methods
were compared and tested on mock galaxy catalogues in 
\citet{2014MNRAS.441.1513O, 2015MNRAS.449.1897O, 2018MNRAS.475..853O,
2018MNRAS.481..324W}. 
This comparison showed that the method used in \citet{2014A&A...566A...1T} 
finds groups reasonably well. However, some individual groups may be
combined or split erroneously. Below we discuss some such 
cases.

The supercluster SCl~A2142 
embeds 14 rich galaxy groups with at least ten member galaxies, 
nine of them in the 
main body of the supercluster at distances from the supercluster centre up to  $11$~\Mpc, 
and five in  the tail.
The supercluster also has
poor groups with less than
ten galaxies, hosting in total 460 galaxies. We consider single galaxies as
the brightest galaxies of faint groups in which other group
members are fainter than the SDSS survey limit \citep{2009A&A...495...37T}.
Data for the groups with at least ten member galaxies
are given in Table~\ref{tab:gr10}.
Masses  of groups in Table~\ref{tab:gr10}
were calculated applying the virial theorem and 
assuming symmetry of galaxy velocity distribution, 
and the Navarro-Frenk-White (NFW)
density profile for galaxy distribution in the plane of the sky.
For a detailed description of how the masses and virial radii of groups were calculated,
we refer to \citet{2014A&A...566A...1T}.

\begin{table*}[ht]
\caption{Data on rich groups in the A2142 supercluster with at least ten member galaxies.}
\begin{tabular}{rrrrrrrrrrr} 
\hline\hline  
(1)&(2)&(3)&(4)&(5)& (6)&(7)&(8)&(9)&(10)&(11) \\      
\hline 
No. & ID&$N_{\mathrm{gal}}$& $\mathrm{R.A.}$ & $\mathrm{Dec.}$ 
&$\mathrm{Dist.}$ &$\mathrm{D_C}$ &$R_{\mathrm{vir}}$ & $L_{\mathrm{tot}}$  
& $M_{\mathrm{dyn}}$ & $D8$  \\
&&&[deg]&[deg]&[$h^{-1}$ Mpc]&[$h^{-1}$ Mpc]&[$h^{-1}$ Mpc]& [$10^{10} h^{-2} L_{\sun}$] 
 & [$10^{12}h^{-1}M_\odot$]  &  \\
\hline
 Main body &&&&&&&&&&\\
\hline                                                    
 1 & 10570 &  27 (31) & 238.53 & 27.47 &  268.6 & 4.4  & 0.53 & 51.3 &   60&  13.7 \\
 2 &  3070 & 212      & 239.52 & 27.32 &  264.6 & 0.0  & 0.88 &382.0 &  907&  20.7 \\
 3 &  4952 &  54 (61) & 239.78 & 26.56 &  260.1 & 3.3  & 0.70 &111.0 &  214&  17.1 \\
 4 & 32074 &  11      & 240.11 & 26.71 &  262.4 & 3.3  & 0.28 & 15.7 &   60&  19.9 \\
 5 & 35107 &  10      & 240.13 & 27.01 &  258.7 & 2.4  & 0.47 & 13.9 &   30&  14.2 \\
 6 & 14960 &  27 (30) & 240.20 & 25.87 &  263.6 & 7.0  & 0.59 & 46.6 &  148&  15.3 \\
 7 & 17779 &  20      & 240.38 & 26.16 &  261.1 & 6.0  & 0.40 & 35.1 &  104&  16.4 \\
 8 &  6885 &  32      & 240.75 & 25.40 &  258.9 & 9.7  & 0.46 & 70.5 &  164&  11.4 \\
 9 & 21183 &  21      & 240.83 & 26.95 &  265.0 & 5.4  & 0.34 & 36.5 &   61&  15.2 \\
\hline                                                    
Tail &&&&&&&&&&\\
\hline                                                    
10 & 20324 &  11 (12) & 240.95 & 24.95 &  259.7 & 12.0 & 0.45 & 17.4 &   78& 10.4 \\
11 & 10818 &  28 (32) & 241.23 & 24.52 &  259.6 & 14.1 & 0.53 & 51.3 &  105&  9.3 \\
12 & 14283 &  19 (25) & 241.90 & 23.21 &  260.3 & 20.9 & 0.40 & 34.4 &   65&  7.3 \\
13 & 10224 &  32      & 242.91 & 21.02 &  254.6 & 32.0 & 0.32 & 60.4 &  233&  6.0 \\
14 & 26895 &  12      & 243.08 & 20.77 &  254.8 & 33.2 & 0.41 & 16.5 &   49&  5.9 \\
\hline
\label{tab:gr10}  
\end{tabular}\\
\tablefoot{                                                                                 
Columns are as follows:
(1): Order number of the group;
(2): ID of the group from \citet{2014A&A...566A...1T} (Gr~3070
correspond to the Abell cluster A2142);
(3): Number of galaxies in the group, $N_{\mathrm{gal}}$. In parenthesis we give
the total number of galaxies in a group, which takes 
into account also galaxies possibly missing from the group
because of fibre collisions;
(4)--(5): Group centre right ascension and declination;
(6): Group centre comoving distance;
(7): Group distance from the  centre of the cluster A2142
(for brevity, clustercentric distance);
(8): Group virial radius;
(9): Group total luminosity;
(10): Dynamical mass of the group assuming the NFW density profile, $M_{\mathrm{dyn}}$;
(11): Luminosity-density field at the location of the group, $D8$, 
in units of the mean density as described in the text.
}
\end{table*}

The SDSS spectroscopic sample is incomplete because of fibre collisions. The 
smallest separation between spectroscopic fibres is 55", and approximately
6\% of the potential targets for spectroscopy are without observed spectra because of this.
We find that in our sample 55 galaxies have a neighbour without measured redshifts 
due to fibre collisions. We 
analysed  the regions around these galaxies with
radius of 55", and searched for galaxies possibly missing from our sample.
We looked for objects in SDSS database classified as galaxies with apparent $r$ magnitudes $r \leq 
17.77$ and photometric redshifts in the same interval
as galaxy or group to which they could belong (within error limits). In this way we have found
31 possible missing member galaxies of the SCl~A2142, 
18 of them in the main body of the supercluster. The number of such galaxies
was the highest in Gr3 from which seven galaxies may be missing due to fibre 
collisions. We used photometric magnitudes of these galaxies when calculating
median colour indexes of groups below.

Absolute magnitudes of galaxies are computed according to the formula
\begin{equation}
M_r = m_r - 25 -5\log_{10}(d_L)-K,
\end{equation} 
where $d_L$ is the luminosity distance in units of $h^{-1}$Mpc and
$K$ is the $k$+$e$-correction. 
The $k$-corrections were calculated with the \mbox{KCORRECT\,(v4\_2)} code
\citep{2007AJ....133..734B} and the evolution corrections have been calibrated 
according to \citet{2003ApJ...592..819B}. 
Details about how the $k$ and $e$-corrections were applied 
can be found in \citet{2014A&A...566A...1T}.
\citet{2014A&A...566A...1T} used  slightly smaller
evolution corrections than  \citet{2003ApJ...592..819B}.
The value of  $M_{\odot} = 4.53$~mag (in $r$-filter). 

At the distance of SCl~A2142, $ \approx 265$~\Mpc, the galaxy
sample is complete
at the absolute magnitude limit $M_r = -19.6$ in units of $\mathrm{mag}+5\log_{10}h$. 
In our study we used the full dataset
of SCl~A2142 to study the group content of the supercluster. 
For comparison of galaxy populations in different groups
we used a magnitude-limited complete sample, excluding galaxies fainter than
the completeness limit, $M_r = -19.6$~mag.  

\subsection{Galaxy populations}
\label{subsec:galpop}

We  obtained data on galaxy properties  from the SDSS DR10 
web page\footnote{\url{http://skyserver.sdss3.org/dr10/en/help/browser/browser.aspx}}.
Galaxy magnitudes and  Petrosian radii were taken from the SDSS 
photometric data.
We calculated galaxy colours as $(g - r)_0 = M_g - M_r$. 
The value $(g - r)_0 = 0.7$ is used to separate red and blue galaxies, where red galaxies 
have $(g - r)_0 \geq 0.7$. 
All magnitudes and colours correspond to  the rest frame at  redshift $z=0$.

Galaxy stellar masses ($M^{\mathrm{*}}$), star formation rates (SFRs), and 
$D_n(4000)$ index of galaxies  
are from the MPA-JHU spectroscopic catalogue \citep{2004ApJ...613..898T, 
2004MNRAS.351.1151B}, from which the various properties of 
galaxies were obtained by fitting SDSS photometry and spectra with
the stellar population synthesis models developed by \citet{2003MNRAS.344.1000B}.
The stellar masses of galaxies were derived as 
described by \citet{2003MNRAS.341...33K}.  The SFRs 
were computed using the photometry and emission lines as described 
by \citet{2004MNRAS.351.1151B} and \citet{2007ApJS..173..267S}. 
The strength of the $D_n(4000)$ break (the ratio of the average flux densities
in the band $4000 - 4100 \angstrom$ and $3850 - 3950 \angstrom$; $D_n(4000)$ index)
is correlated with the time passed from the most recent star formation event 
and is defined as in \citet{1999ApJ...527...54B}. 
The $D_n(4000)$ index characterises star formation histories of galaxies and 
can be understood as an indicator for luminosity-weighted age.

The $D_n(4000)$ index was used to separate quiescent and star-forming
galaxies. The higher the value of the $D_n(4000)$ index, the older
are stellar populations in a galaxy. 
We applied values for  quiescent galaxies with old stellar populations 
as having  $D_n(4000) \geq 1.55$, 70\% of all galaxies (73\% in the  main body
and 63\% in the tail). 
Star-forming galaxies with young stellar 
populations have $D_n(4000) < 1.55$. This  
limit was also used by \citet{2003MNRAS.341...54K} and \citet{2017A&A...605A...4H} 
to separate young and old galaxies 
in the SDSS survey. 

Stellar ages are from the Portsmouth group \citep{2009MNRAS.394L.107M}.
We divided galaxies into old and young stellar populations by stellar age
using the age limit $t = 3$~Gyr.
We found that 61\% of galaxies in the supercluster 
have old stellar populations with stellar ages $t \geq 3$~Gyr 
(in supercluster  main body 63\%, and in the tail 56\%). 
The  $\log \mathrm{SFR} \leq -0.5$ 
corresponds to quiescent galaxies (88\% of galaxies). 
Actively star-forming galaxies are characterised by 
$\log \mathrm{SFR} > -0.5$; this limit was also applied in   
\citet{2014A&A...562A..87E}.

Some star-forming galaxies have red colours and are known as red star-forming
galaxies. Also, there are galaxies with low SFRs
but also
low values of $D_n(4000)$ index, which suggest that they may be recently quenched.
We define recently quenched galaxies as those
with $D_n(4000) \leq 1.55$ and  $\log \mathrm{SFR} < -0.5$ (31 galaxies, 0.03\%
of all galaxies).
Red, high SFR galaxies are defined as galaxies with 
$g - r \geq 0.7$ and $\log \mathrm{SFR} \geq -0.5$ (138 galaxies, 13\% of galaxies). 

To identify galaxies with active nuclei (AGNs) we used 
the spectral classification  by \citet{2004MNRAS.351.1151B}
based on emission-line ratios 
[\ion{O}{iii}]$/$H$\beta$ and [\ion{N}{ii}]$/$H$\alpha$.
We find that SCl~A2142 embeds 17 AGNs.
According to the VLA FIRST survey data at 1.4 GHz 
\footnote{\url{http://sundog.stsci.edu/}. }
three of them are very low flux radio sources.
We also identify  FR II radio galaxy (J155852.66 +262618.9) in the group Gr3
using the catalogue of FR II radio galaxies by \citet{2017A&A...601A..81C}.

We show the sky distribution of galaxies in the supercluster
in Fig.~\ref{fig:radec}. Figure~\ref{fig:radec} shows that most rich groups  
in the supercluster main body lie along the supercluster axis
(on-axis groups). 
Only groups Gr1 and Gr9
lie farther away from the axis (off-axis groups, we call them 
according to their order number from Table~\ref{tab:gr10}).
For details about SCl~2142 we refer to \citet{2015A&A...580A..69E, 2018A&A...610A..82E}.

\begin{figure}[ht]
\centering
\resizebox{0.45\textwidth}{!}{\includegraphics[angle=0]{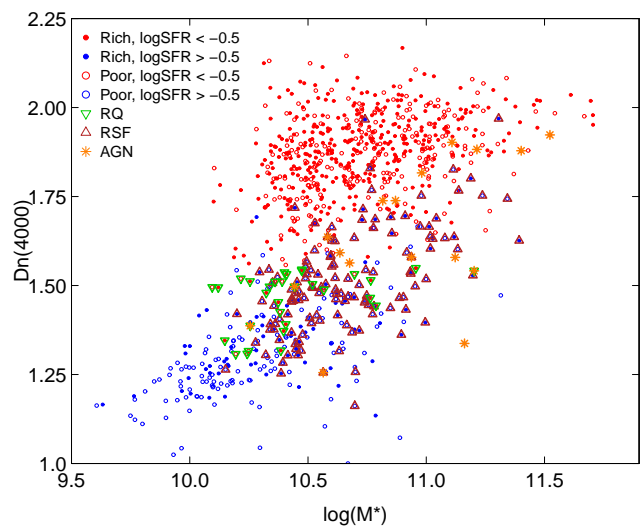}}
\caption{
$D_n(4000)$ index vs. stellar mass for SCl~A2142.
Red circles indicate low star formation rate galaxies with 
$\log \mathrm{SFR} < -0.5$, and blue circles shows high star formation rate
galaxies with $\log \mathrm{SFR} \geq -0.5$. 
Filled circles correspond to rich groups with $N_\mathrm{gal} \geq 10$,
and empty circles show galaxies in poor groups with $N_\mathrm{gal} < 10$.
The red triangles indicate red, high SFR galaxies  defined as 
$g - r \geq 0.7$, and $\log \mathrm{SFR} \geq -0.5$. 
Other notations are as in 
Fig.~\ref{fig:radec}.
}
\label{fig:a2142d4m}
\end{figure}

\begin{figure}[ht]
\centering
\resizebox{0.45\textwidth}{!}{\includegraphics[angle=0]{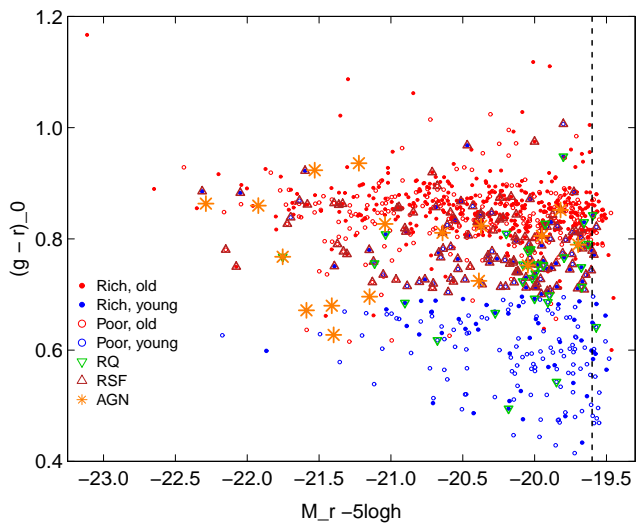}}
\caption{
Colour-magnitude diagram $(g - r)_0$ vs. $M_r$ for galaxies in SCl~A2142.
Filled circles correspond to galaxies in rich groups,
and empty circles show galaxies in poor groups
(including single galaxies). 
The red circles correspond to the galaxies with old stellar populations
($D_n(4000) \geq 1.55$) and the blue circles to the galaxies having young 
stellar populations with $D_n(4000) < 1.55$. The green triangles show 
recently quenched  galaxies 
with $D_n(4000) \leq 1.55$ and star formation rate $\log \mathrm{SFR} < -0.5$.
The red triangles indicate red, high SFR galaxies  defined as 
$g - r \geq 0.7$, and $\log \mathrm{SFR} \geq -0.5$.
Orange stars represent AGNs.
The dashed line shows the completeness limit $M_r = -19.6$~$+5\log_{10}h$.
}
\label{fig:a2142cmr}
\end{figure}

To characterise galaxies with different stellar populations
we plot in Fig.~\ref{fig:a2142d4m} $D_n(4000)$ index versus stellar mass of galaxies
in the supercluster.
The colour-magnitude diagram for SCl~A2142 is shown in Fig.~\ref{fig:a2142cmr}.
We plot galaxies with old and young stellar populations, recently quenched
galaxies, and red, star-forming galaxies with different symbols.
Here we also show the completeness limit in luminosity,
$M_r = -19.6$~$+5\log_{10}h$~mag. 

\section{Methods}
\label{sect:medthods} 

\subsection{Dynamical state of groups: phase space diagram}
\label{sect:ppsm} 

The PPS diagram shows 
line-of-sight velocities of galaxies 
with respect  to the cluster mean velocity (velocity offset)
versus projected clustercentric distance. In this paper clustercentric
distance 
$D_{\mathrm{C}}$ 
is defined as the projected distance from the centre of the
cluster A2142 (actually  superclustercentric distance; we use the term
clustercentric distance for brevity).
In the phase space diagram galaxies with 
different accretion histories populate different areas 
\citep{2013MNRAS.431.2307O, 2014ApJ...796...65M, 2015ApJ...806..101H, 2015MNRAS.448.1715J,
2017MNRAS.467.4410A, 2017ApJ...838..148P, 2017ApJ...838...81Y, 2017ApJ...843..128R}. 
Galaxies with early infall times (infall at  redshifts $z > 1$)
are located at small clustercentric distances
in the virialised region. Recently infallen or still infalling 
galaxies lie at large
clustercentric distances and typically have velocity offsets around zero or
with negative values (non-virialised region) with a small scatter
\citep{2007MNRAS.376.1577D, 2015ApJ...806..101H}.  

Galaxies with
large clustercentric distances and positive velocity offsets
may also be backsplash galaxies -- galaxies
which have passed cluster centre already at least once and now are orbiting
at large clustercentric distances \citep{2015ApJ...806..101H}.
For backsplash galaxies see also \citet[][and references
therein]{2013MNRAS.430.3017B}. 

The analysis of the PPS diagram has become an important tool to
study galaxy populations in galaxy groups and clusters,
to analyse dynamical properties of galaxies in clusters and outskirts and
to compare galaxy populations in virialised and non-virialised regions
of the clusters \citep{2015ApJ...806..101H, 2015MNRAS.448.1715J,
2017MNRAS.467.4410A, 2017ApJ...838..148P, 2017ApJ...838...81Y, 2017ApJ...843..128R,
2017MNRAS.471..182W}. 

We used projected phase space diagram for the  main body of the
A2142 supercluster to analyse the 
dynamical state of different galaxy groups in the supercluster,
and their possible infall into the main cluster of the supercluster.
We also used it to analyse the distribution of various
galaxy populations in groups.

\subsection{Substructure: normal mixture modelling}
\label{sect:mclust} 

Multi-dimensional normal mixture modelling was used
to  search for substructure in galaxy groups. For this purpose 
we apply {\it mclust} 
package for classification and clustering
\citep{fraley2006} from {\it R}, an open-source free statistical environment 
developed under the GNU GPL \citep[][\texttt{http://www.r-project.org}]{ig96}.
This package is based on the analysis of a finite mixture of 
distributions, 
in which each mixture component is taken to correspond to a 
different sub-group of the cluster.
By default, {\it mclust} analyses 14 different models with
up to nine components.
The best solution for components found by {\it mclust}
is chosen using the Bayesian information criterion (BIC),
the best solution has the highest value of BIC
among all models and number of components considered by {\it mclust}. 
In {\it mclust} for every galaxy the probability to belong to any
of the components is calculated, and the uncertainty of classification is defined 
as one minus the highest probability of a galaxy to belong to a component.
Galaxies are assigned to components based on the highest probability
calculated by {\it mclust}.  

This analysis was applied only to groups that had  more than 12 galaxies; 
for groups with a smaller
number of galaxies the results of {\it mclust} are less reliable,
as shown by \citet{2013MNRAS.434..784R}. 
As an input for {\it mclust} we used 
the sky coordinates and velocity of the group member galaxies. 
The values of velocities were scaled to make them of the same order as the values
of coordinates.  
In \citet{2012A&A...540A.123E} and \citet{2018A&A...610A..82E}
this method  was applied to analyse the substructure in galaxy groups and
clusters.
The same methodology was also used by \citet{2016A&A...588A..14T, 2017A&A...602A.100T}
to detect sub-components of friends-of-friends galaxy groups in order to  
refine group determination, and to find merging galaxy groups.

\subsection{Magnitude gap between the brightest galaxies in groups}
\label{sect:mgap} 
Magnitude gap between the brightest galaxies in groups
is one indicator of the formation history of the groups and clusters,
and their dynamical state 
\citep[][and references therein]{2017ApJ...845...45K,
2017MNRAS.472.3246M, 2018MNRAS.474..866V}.
Large magnitude gaps of the brightest galaxies in clusters
suggests that such clusters may have formed earlier than groups
and clusters with small magnitude gaps. They may also 
have different recent accretion history, groups or clusters with 
large magnitude gaps having larger time since the
last major merger in them.

\subsection{Orientations: group position angles, brightest galaxies, and supercluster 
axis }
\label{sect:ori} 

To determine the position angle of the supercluster axis and individual
groups, we approximated their shape with the ellipse and found the position angle of its
major axis. 
Details of this analysis were presented in \citet{2018A&A...610A..82E}.
The position angle of the supercluster axis in Fig.~\ref{fig:radec} was
determined by the distribution of rich galaxy groups
in the supercluster \citep[see also ][]{2018A&A...610A..82E}. 
The position angle of the supercluster axis is $63 \pm 1^{\circ}$,
measured counterclockwise from west.
Below we compare the orientation of the supercluster axis
and galaxy groups, and check for the
alignments between the brightest group galaxies, group and supercluster
axes.
 

\begin{table*}[ht]
\caption{Parameters of galaxies in groups.}
\begin{tabular}{rrrrrrrrrrrrr} 
\hline\hline  
(1)&(2)&(3)&(4)&(5)& (6)&(7) &(8)&(9)& (10)&(11)&(12)&(13)\\      
\hline 
No. & ID & \multicolumn{2}{c|}{$log M^{\mathrm{*}}$}   
& \multicolumn{2}{c}{$D_n(4000)$} &\multicolumn{2}{c}{$(g - r)_0$}  
& \multicolumn{2}{c}{$SFR$} & \multicolumn{2}{c}{$t$}& $|\Delta M_{12}|$\\
  &    & ${\mathrm{median}}$   & $F_{\mathrm{h}}$
& ${\mathrm{median}}$ & $F_{\mathrm{1.55}}$ & ${\mathrm{median}}$ & $F_{\mathrm{red}}$ & 
${\mathrm{median}}$   & $F_{\mathrm{p}}$ & ${\mathrm{median}}$ & $F_{\mathrm{old}}$&$mag$\\
\hline
\multicolumn{2}{l|}{ Main body} & &  &  &  &   &  & & &   & &\\
\hline
 1 & 10570   & 10.68  & 0.97 & 1.67  & 0.65 & 0.84 (0.85) & 0.74 (0.74) & -0.44  & 0.48 & 3.0  & 0.52 &0.04\\
 2 &  3070   & 10.69  & 0.81 & 1.87  & 0.88 & 0.86        & 0.92        & -1.29  & 0.87 & 8.3  & 0.77 &1.24\\
 3 &  4952   & 10.58  & 0.96 & 1.81  & 0.73 & 0.83 (0.83) & 0.85 (0.82) & -0.98  & 0.59 & 4.4  & 0.59 &0.33\\
 4 & 32074   & 10.47  & 1.0  & 1.84  & 0.74 & 0.82        & 0.91        & -1.20  & 0.73 & 5.5  & 0.73 &0.17\\
 5 & 35107   & 10.56  & 1.0  & 1.82  & 0.90 & 0.81        & 0.90        & -1.15  & 0.90 & 4.5  & 0.90 &0.18\\
 6 & 14960   & 10.68  & 0.85 & 1.87  & 0.78 & 0.83 (0.84) & 0.81 (0.83) & -1.21  & 0.70 & 5.0  & 0.63 &0.58\\
 7 & 17779   & 10.58  & 1.0  & 1.75  & 0.63 & 0.83        & 0.85        & -1.04  & 0.60 & 4.6  & 0.63 &0.0\\
 8 &  6885   & 10.61  & 0.97 & 1.90  & 0.91 & 0.85        & 0.94        & -1.35  & 0.93 & 8.4  & 0.81 &0.39\\
 9 & 21183   & 10.67  & 0.86 & 1.65  & 0.67 & 0.78        & 0.76        & -0.72  & 0.62 & 3.8  & 0.57 &0.04\\
\hline                                                                                                  
Tail & & &  &  &  &   &  & & &   & &\\                                                                  
\hline                                                                                                  
 10 & 20324   & 10.44 & 0.73 & 1.85 & 0.82  & 0.84 (0.84) & 0.91 (0.92) & -1.24  & 0.67 & 5.5  & 0.73 &0.72\\
 11 & 10818   & 10.72 & 0.93 & 1.89 & 0.79  & 0.86 (0.86) & 0.86 (0.88) & -1.05  & 0.75 & 5.8  & 0.68 &0.64\\
 12 & 14283   & 10.52 & 0.89 & 1.80 & 0.68  & 0.83 (0.86) & 0.85 (0.88) & -1.15  & 0.79 & 5.8  & 0.74 &0.01\\
 13 & 10224   & 10.82 & 0.95 & 1.90 & 0.81  & 0.84        & 0.90        & -1.25  & 0.81 & 6.3  & 0.71 &0.28\\
 14 & 26895   & 10.63 & 0.83 & 1.76 & 0.82  & 0.82        & 0.92        & -0.89  & 0.75 & 2.9  & 0.50 &0.20\\
\hline                                                                  
 P(main) &    & 10.59 & 0.85& 1.68  & 0.60 & 0.80  (0.81) & 0.72 (0.73) & -0.78  & 0.57 & 3.3  & 0.53 &\\
 P(tail) &    & 10.54 & 0.84& 1.67  & 0.55 & 0.80  (0.80) & 0.69 (0.69) & -0.56  & 0.51 & 3.0  & 0.50 & \\
\hline                                                 
\label{tab:galpara}  
\end{tabular}\\
\tablefoot{                                                                                 
Columns are as follows:
(1): Order number of the group. P denotes galaxies in poor groups. 
Main mark galaxies in the main
body of the supercluster up to clustercentric distances $11.5$~\Mpc,
and tail - galaxies with clustercentric distances larger than $11.5$~\Mpc;
(2): ID of the group;
(3-12): Median values of galaxy parameters,
and percentages of galaxies of high stellar mass, of old stellar populations according
to the $D_n(4000)$ index,
red galaxies, passive galaxies, and galaxies with old stellar ages
in a group.
Values in parenthesis show $(g - r)_0$ 
corresponding parameters calculated including colours of possible group members missing
because of fibre collisions. 
(13) Magnitude gap between the two brightest galaxies in a group.
For A2142 we give this for the BCG1 and BCG3 \citep{2018A&A...610A..82E}.
}
\end{table*}

In the next two sections  we analyse the galaxy content of galaxy groups in SCl~A2142,
and their location in the PPS diagram. The properties of individual
groups are given in Appendix~\ref{sect:gr14}.

\section{Galaxy populations in groups }
\label{sect:galpop} 

\begin{figure}[ht]
\centering
\resizebox{0.47\textwidth}{!}{\includegraphics[angle=0]{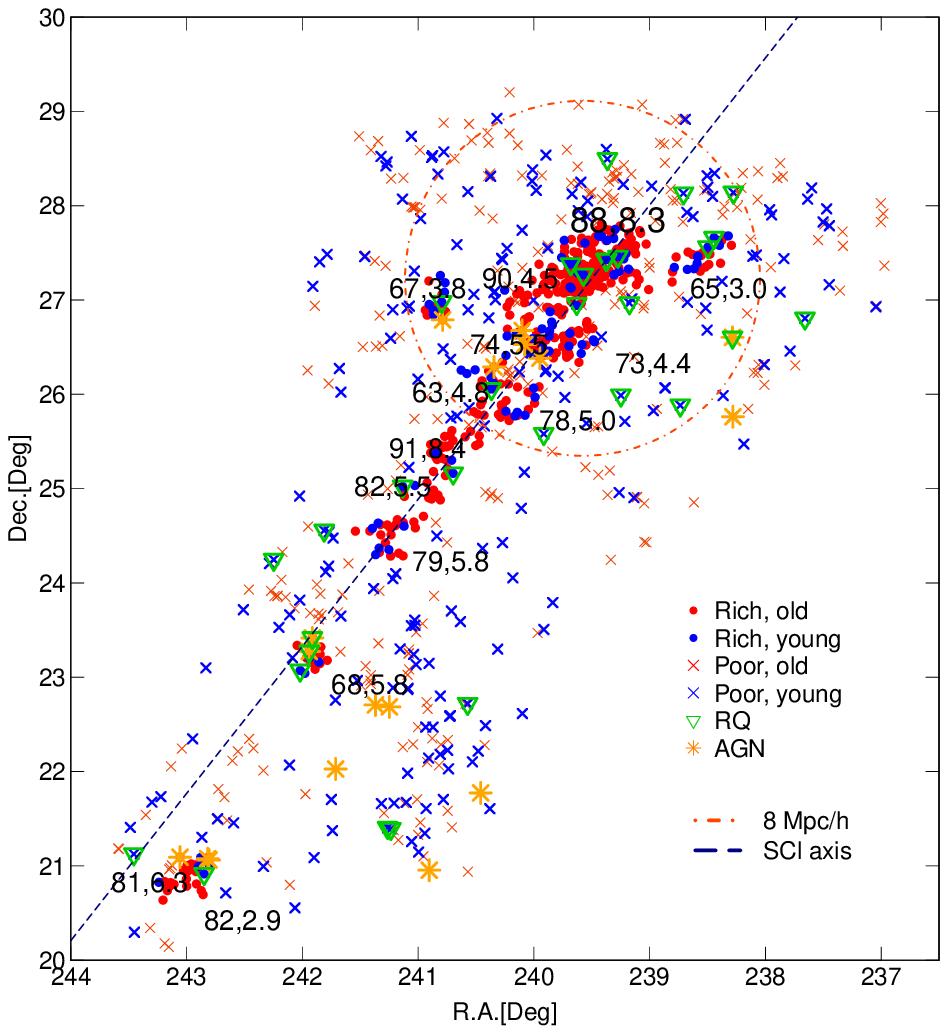}}
\caption{
Sky distribution of galaxies in SCl~A2142. 
The red symbols show galaxies with old stellar populations
($D_n(4000) \geq 1.55$), and the blue symbols denote galaxies with young 
stellar populations ($D_n(4000) < 1.55$).
The first number shows the percentage of galaxies with old stellar
populations divided by $D_n(4000)$ index
in rich groups. The second number shows 
the median age of galaxy populations in a group (in Gyr, Table~\ref{tab:galpara}).
Galaxies in rich groups  are denoted with filled circles.
The crosses denote galaxies in poor groups with $2 - 9$ galaxies,
and single galaxies. 
The green triangles show recently quenched  galaxies, and  
orange stars indicate the positions of AGNs.
The orange circle has a radius 
of about $8$~\Mpc\ (in sky projection).
The dashed line shows the supercluster axis.
}
\label{fig:radecfro}
\end{figure}

We analysed the full distributions of galaxy parameters
in groups, and their median values.
Our aim is to compare the galaxy content of groups 
in various locations in the supercluster, to see whether groups at different
clustercentric radii in the supercluster main body and tail,
and groups on the supercluster axis and off-axis groups
have similar or different galaxy populations.

We present
median values of stellar masses (${\mathrm{log}} M^{\mathrm{*}}$), $D_n(4000)$ indexes, 
colours ($(g - r)_0$), SFRs ($SFR$), and stellar ages ($t$)
of galaxies in rich groups in Table~\ref{tab:galpara}. 
Table~\ref{tab:galpara} gives also the percentages of galaxies with old stellar populations
according to the $D_n(4000)$ index, $F_{\mathrm{1.55}}$, and stellar age, $F_{\mathrm{old}}$,
the percentage of red galaxies, $F_{\mathrm{red}}$, passive galaxies,  $F_{\mathrm{p}}$,
and the percentages of high stellar mass galaxies with stellar
masses ${\mathrm{log}} M^{\mathrm{*}} \geq 10.25$ 
\citep[$F_{\mathrm{h}}$, see][for the choice of this limit]{2014A&A...562A..87E}.
For comparison we show parameters of galaxies in poor groups (denoted as P).
For groups with galaxies missing due to the fibre collisions 
we give in Table~\ref{tab:galpara} also the median value of colour index $(g - r)_0$ and 
percentages of red galaxies 
calculated including colours of possible group members. Table~\ref{tab:galpara}
shows that they only have minor effect.  In Fig.~\ref{fig:radecfro} we show the 
sky distribution of galaxies from different populations. For each rich group
we give the fraction of 
galaxies with old stellar populations according to the $D_n(4000)$ index,
and the median stellar age (Table~\ref{tab:galpara}).
Figure~\ref{fig:radecfro} gives a general picture of 
galaxy populations in groups in the supercluster.

\begin{figure*}[ht]
\centering
\resizebox{0.30\textwidth}{!}{\includegraphics[angle=0]{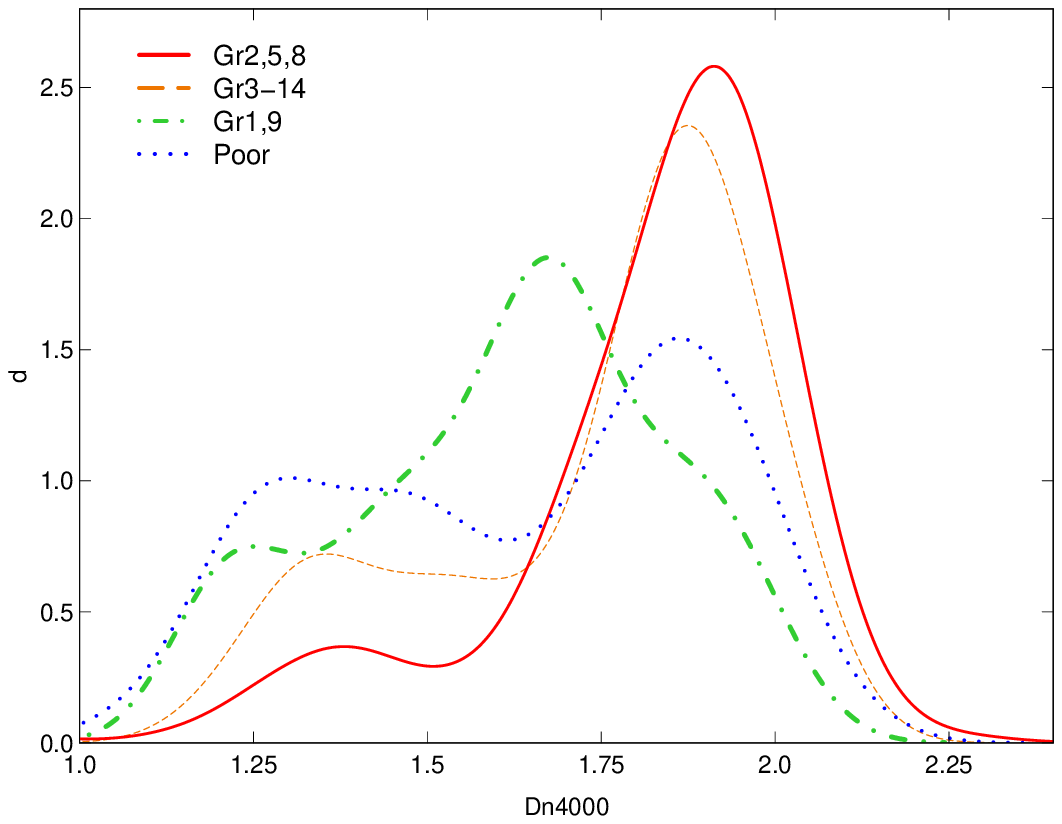}}
\resizebox{0.30\textwidth}{!}{\includegraphics[angle=0]{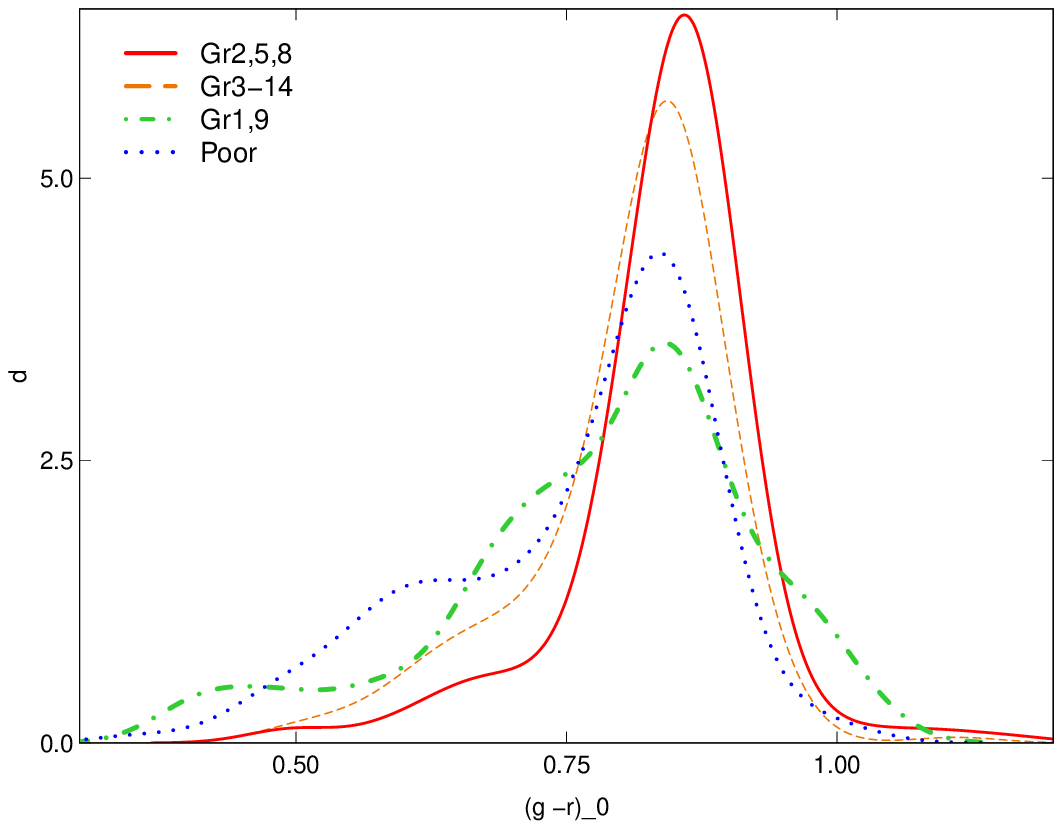}}
\resizebox{0.30\textwidth}{!}{\includegraphics[angle=0]{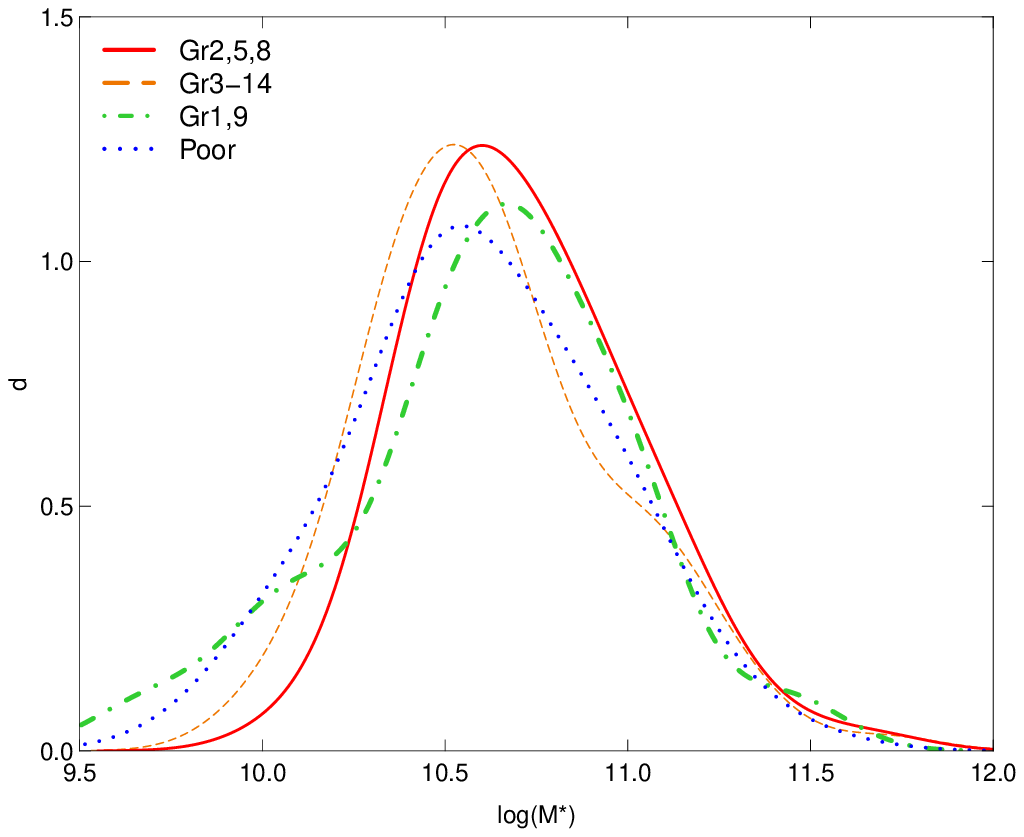}}
\caption{
Probability density distributions of $D_n(4000)$ index (left panel),
$(g - r)_0$ colour index (middle panel), and stellar masses for galaxies 
(right panel) in groups
2, 5, and 8 (red solid line), groups 3, 4, 6, 7, and 10 - 14 
(notation Gr3-14, orange dashed line), 
in groups 1 and 9 (green dot-dashed line), and in poor groups
(blue dotted line). Group ID numbers are given in Table~\ref{tab:gr10}.
}
\label{fig:galpropdistr}
\end{figure*}

Table~\ref{tab:galpara} and Fig.~\ref{fig:radecfro}  
show that groups with the oldest 
stellar populations and  lowest star formation rates reside in
the cluster A2142 (Gr2), in Gr5 close to A2142, 
and in the group Gr8 on the supercluster axis at the edge of the main
body of the supercluster where  
the supercluster tail begins. Approximately $90$~\%
of galaxies in these groups have old stellar populations
according to $D_n(4000)$ index. Also, according to the stellar age 
galaxies in A2142 cluster and Gr8 have
median stellar populations older than $8$~Gyr.
Galaxy populations in the cluster A2142 and in its infalling subclusters 
were analysed in detail in \citet{2018A&A...610A..82E}. This analysis shows that
galaxies with the oldest stellar populations and with the lowest star formation rates
lie in the central, virialised region of A2142, and in one infalling
subcluster, called as M1 by \citet{2018A&A...610A..82E}. We compare
M1 and Gr5 in Appendix~\ref{sect:gr14}. 

Groups on the supercluster axis have median ages of stellar populations
approximately $4 - 7$~Gyr, and the fraction of galaxies with old
stellar populations $70 - 80$~\% (except Gr7 with 63\% of old
galaxies). 
In the supercluster main body groups with galaxies having the youngest stellar
ages  ($t < 4$~Gyr) and with the 
lowest percentages of galaxies with old stellar
populations are Gr1 and Gr9 away from supercluster axis.
The galaxy properties in these off-axis groups are more similar to 
those in poor groups which embed galaxies with younger stellar populations
than rich groups in average.
Groups with younger galaxy populations have also larger percentages of blue galaxies.

We used the Kolmogorov-Smirnov (KS) test to find the statistical significance 
of the differences in galaxy populations for galaxies in groups.
We considered that the differences between distributions are highly significant 
if the $p$ - value (the estimated probability of rejecting the hypothesis
that distributions are statistically similar) is $p \leq 0.01$.

The statistical comparison of galaxy content of groups by
individual groups is complicated due to small number of galaxies 
in groups with 12 or fewer member galaxies (four groups). 
In this case the results of the tests
are not reliable, as shown, for example, in \citet{2013MNRAS.434..784R}.
Therefore, after initial comparison of galaxy content of groups
we combined groups as follows. 
Groups 2, 5, and 8 can be taken together as the ones with the oldest stellar populations.
The  comparison of galaxy content of other groups shows that
groups on the supercluster axis
in the supercluster main body  and in the tail do not have
statistically significant differences between their galaxy populations, therefore
groups 3, 4, 6, and 7 from the supercluster main body, together with groups
10 - 14 from the supercluster tail (notation Gr3-14 from group ID-s, 3 - 14) form
an on-axis group class. Off-axis groups Gr1 and Gr9 form a separate class
(off-axis groups). 
Poor groups 
in the supercluster main body and tail have statistically similar galaxy populations and
they can be taken together as one class of groups.
In Fig.~\ref{fig:galpropdistr} 
we show the probability density distributions of $D_n(4000)$ index,
$(g - r)_0$ colour index, and stellar masses ${\mathrm{log}} M^{\mathrm{*}}$ 
for galaxies in groups from these divisions. 

The KS test shows that the differences between
$D_n(4000)$ indexes and colour indexes in 
the cluster A2142 and in Gr8 and Gr5 taken together, and in 
all other groups are significant at a very high 
level, with $p < 0.01$. 
Also, galaxy populations in groups on the supercluster
axis and in off-axis groups are different at very high 
significance level, with $p < 0.01$. 

Galaxies in the cluster A2142 and in Gr8 and Gr5
have higher stellar masses than galaxies in other groups
(or in poor groups), with the KS test $p$ - value $p \approx 0.01$.
The stellar masses of galaxies in groups on the supercluster
axis and in off-axis groups 
are statistically similar, with $p > 0.1$.
This shows that galaxies with the same stellar mass 
in different 
groups have different star formation  histories.

In Table~\ref{tab:galpara} we also present the magnitude gaps of the brightest 
galaxies in SCl~A2142 groups  in $r$-band.
The Table shows that the magnitude gap is larger than one magnitude
in the cluster A2142 only. This suggests that 
the time since the last major merger in A2142 may be at least $4$~Gyrs  
\citep{2013ApJ...777..154D, 2017MNRAS.472.3246M}.
The magnitude gaps between the brightest galaxies in the cluster A2142 were 
analysed in more detail in \citet{2018A&A...610A..82E}. 
In other groups the magnitude gap is less than one magnitude, which indicates that
groups are dynamically young. 

\section{Phase space diagram and clustercentric distances}
\label{sect:pps} 

\begin{figure}[ht]
\centering
\resizebox{0.47\textwidth}{!}{\includegraphics[angle=0]{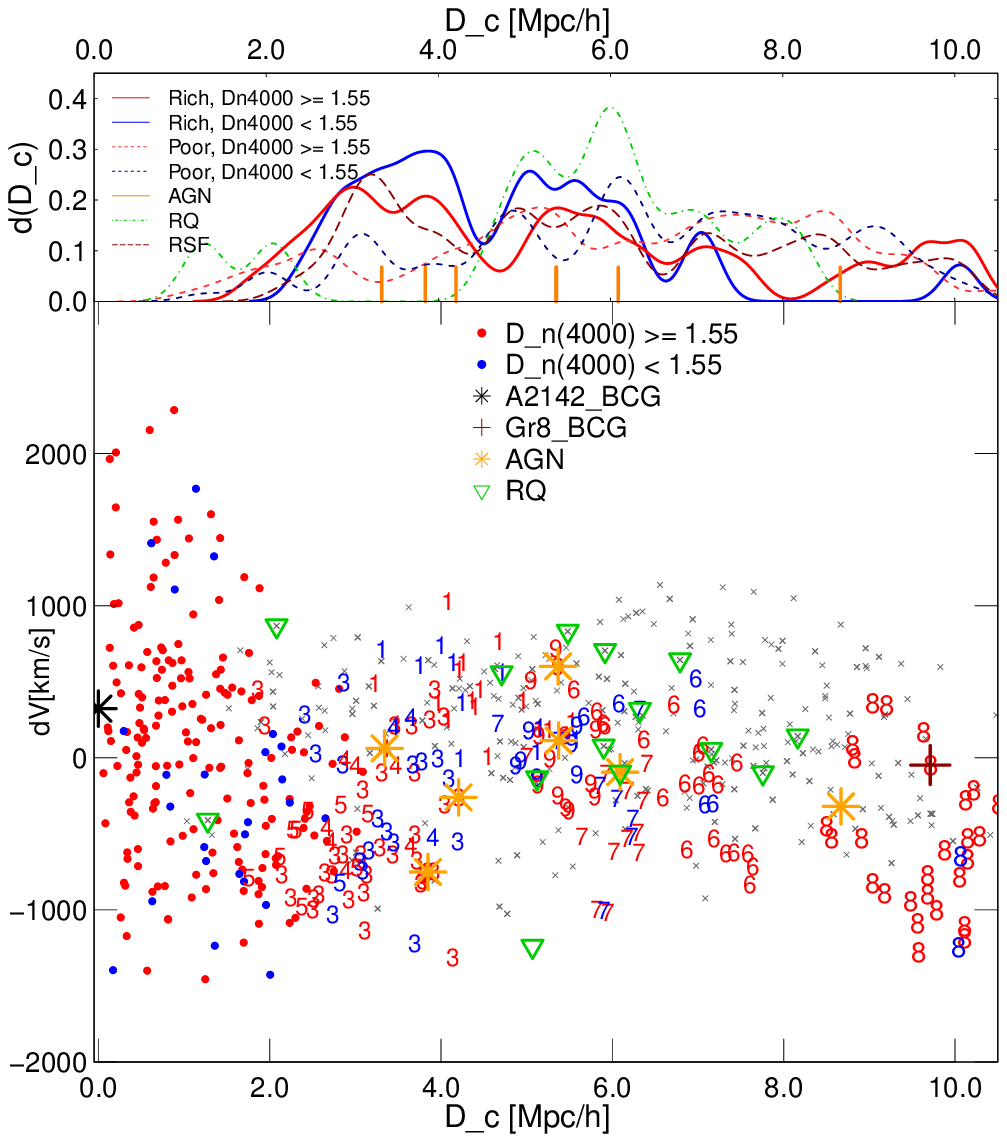}}
\caption{
Upper panel: Distribution of clustercentric distances 
for galaxies with young and old stellar populations in rich groups up to 
clustercentric distances $D_c \leq 10.5$~\Mpc\ 
(solid lines, $D_n(4000) \geq 1.55$), red line, and ($D_n(4000) < 1.55$, blue line),
and in poor groups (dashed red and blue lines).
Galaxies from the cluster A2142 are not included when calculating distributions.
Green and dark red dashed lines show distribution of clustercentric distances 
of recently quenched and red star forming  galaxies, respectively (see text
for definition). 
Orange ticks show clustercentric distances of AGNs.
Lower panel: Velocity of galaxies with respect to the cluster mean velocity vs. 
projected clustercentric distance for the supercluster A2142 main body up to 
clustercentric distances $D_c \leq 10.5$~\Mpc\ (PPS diagram).
The red symbols show galaxies with old stellar populations
($D_n(4000) \geq 1.55$), and the blue symbols denote galaxies with young 
stellar populations ($D_n(4000) < 1.55$).
Filled circles correspond to galaxies in the cluster A2142.
Galaxies in other rich groups are plotted with  
the number which is the order number of a group in Table~\ref{tab:gr10}.
Grey crosses indicate galaxies in poor groups with $2 - 9$ galaxies,
and single galaxies. Triangles show recently quenched galaxies, 
as in Fig.~\ref{fig:radec}.
}
\label{fig:a2142cdv8}
\end{figure}

Figure~\ref{fig:a2142cdv8} presents the PPS diagram (velocity offset versus
the projected clustercentric distance) for galaxies in different
groups up to projected clustercentric distances $10.5$~\Mpc\ from the supercluster
centre. Numbers show order numbers of groups
from Table~\ref{tab:gr10} where galaxies lie, and crosses indicate the positions
of galaxies from poor groups or single galaxies. 
We also plot the location of recently quenched galaxies.
Red star forming galaxies are mostly located in rich groups. We have not indicated them 
in Fig.~\ref{fig:a2142cdv8} to avoid  overcrowding figure.
In the upper panel of Fig.~\ref{fig:a2142cdv8} 
we show the probability density distributions of clustercentric distances
of galaxies in rich and poor groups from various populations. The members
of the cluster A2142 were not included in these calculations
since their distribution was analysed in detail in \citet{2018A&A...610A..82E}.
Distributions are normalised, so that each integrates to 1.
We do not show error limits in the probability density distributions
in upper panel of Fig.~\ref{fig:a2142cdv8} since errors are sensitive to binning
the data. We used full data (the integral distributions)
to apply the Kolmogorov-Smirnov test for calculating the statistical significance 
of the differences between various distributions  
\citep[see also comments on errors in ][]{2008ApJ...685...83E}.

Galaxies in Fig.~\ref{fig:a2142cdv8} at projected  clustercentric distances 
$D_c < 2$~\Mpc\ belong mostly to the cluster A2142,
with virial radius of $0.9$~\Mpc\ and infall region of subclusters
at radius of about $2$~\Mpc\ \citep{2018A&A...610A..82E}. High peculiar velocities of 
galaxies indicate  the Fingers-of-God effect. At  higher clustercentric distances
galaxies have mostly negative velocity offset in respect to the supercluster centre.
They lie in  the infall region of A2142 cluster.  
Thus all galaxies  in the supercluster core may be falling into the main cluster.
Of course, one must be cautious when interpreting observational data since
we only have redshift information and sky coordinates of the galaxies.
Also, as we show below, Gr8 is probably not falling into the main cluster.

Figure~\ref{fig:a2142cdv8} reveals several interesting details about
galaxies in the PPS diagram. Upper panel shows that the distributions
of clustercentric distances of galaxies in rich groups
have wide maxima at  distances of about $2 - 4$ and $4 - 6$~\Mpc. 
Comparison with  Fig.~\ref{fig:radec} shows  that the maximum at
$4$~\Mpc\ is formed by galaxies in groups Gr1 (1 in Table~\ref{tab:gr10}), 
Gr3, and Gr4.
Together with A2142 they form the highest density core of the supercluster
\citep{2015A&A...580A..69E}. 
The maximum in clustercentric distance distribution at $5 - 6$~\Mpc\ is 
formed by groups Gr6, Gr7, and Gr9. 

In Fig.~\ref{fig:a2142cdv8} the distributions of clustercentric distances
of galaxies in rich groups from old and young populations are shifted. 
In the distance interval up to $4$~\Mpc\ galaxies with old
stellar populations lie
closer to the main cluster than galaxies with young stellar populations. 
Galaxies with old stellar populations have a maximum in the clustercentric 
distance distribution at approximately $3$~\Mpc, while galaxies with 
young stellar populations have this maximum at $4$~\Mpc.
The KS test shows
that distance distributions of galaxies from different populations
are statistically different with
a high significance ($p$-values $p < 0.01$). 

In poor groups 
the distribution of clustercentric distances of 
galaxies with young stellar populations
has a maximum around $6$~\Mpc. 
Figures~\ref{fig:radec} and \ref{fig:a2142cdv8}
show that recently quenched galaxies mostly follow the distribution
of galaxies in poor groups. Only 30\% of these galaxies lie in rich groups.
Of 13 recently quenched galaxies 
eight reside at clustercentric distances  $5-7$~\Mpc.
At this distance interval 
star-forming galaxies with young stellar populations
and recently quenched galaxies are spread over the whole 
main body of the supercluster. They are neither preferentially located along the supercluster axis nor in filaments.  

In the supercluster core up to clustercentric distances of 
$3 - 4$~\Mpc\ AGNs lie in  a small region of the sky plane,  
in groups 3 and 4. Three of them are classified as AGNs in the SDSS database, and
one is FRII radio galaxy.  Other AGNs in the supercluster core lie at 
clustercentric distance interval $5 - 6$~\Mpc, in groups 7 and 9,
and two of them are single galaxies. 
In the supercluster tail AGNs lie on groups 12, 13, and 14, and among poor groups
in a lower density part of the tail. Their distribution follows the 
distribution of recently quenched galaxies.

Among red star-forming galaxies 60\% lie in rich groups.
We do not show them on the lower
panel of Fig.~\ref{fig:a2142cdv8}, but present the distribution
of their clustercentric distances in the upper panel of the figure.
This distribution approximately
follows that of galaxies in groups. The KS test shows that the distributions
of clustercentric distances of red star forming galaxies, and
galaxies in rich and poor groups are statistically similar, with $p$-value
$p > 0.12$.
This agrees with earlier findings that red star forming galaxies
in superclusters 
can be found in groups of various richness, they may also
be central galaxies of groups
\citep[][and references therein]{2005A&A...443..435W, 2008ApJ...685...83E,
2014A&A...562A..87E, 2018MNRAS.476.5284E}.

\section{Discussion}
\label{sect:discussion} 

\subsection{Collapse of the supercluster core}
\label{sect:collapse} 
  
\begin{figure}[ht]
\centering
\resizebox{0.45\textwidth}{!}{\includegraphics[angle=0]{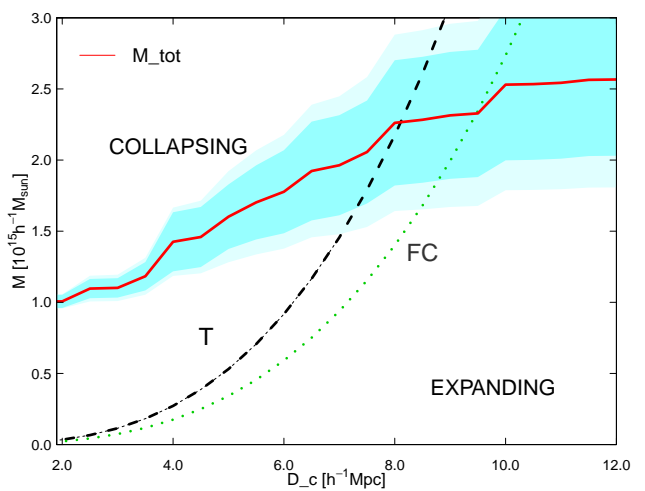}}
\caption{
Mass versus clustercentric distance 
for the main body of SCl~2142. Red solid line shows the total mass within a radius $D_c$, 
and shaded areas show error limits as explained in the text.
Black and green dotted lines show turnaround (T) and future collapse (FC) mass 
($M_{\mathrm{T}}(R)$ and $M_{\mathrm{FC}}(R)$) 
versus radius of a sphere $R$ in a spherical collapse model (Appendix~\ref{sect:sph}). 
}
\label{fig:massradius}
\end{figure}

\citet{2015A&A...580A..69E} and \citet{2015A&A...581A.135G}  
applied the spherical collapse model to analyse 
the dynamical state of the main body of SCl~2142. The spherical collapse
model describes the evolution of a 
spherically symmetric perturbation in an expanding universe. 
We summarise this model shortly in Appendix~\ref{sect:sph}.
In this model the dynamics of a collapsing shell is determined 
by the mass in its interior. 
\citet{2015A&A...580A..69E} and \citet{2015A&A...581A.135G}
defined a high-density core of the supercluster and its size 
using density contrasts in the luminosity-density field. 
They determined the dynamical mass embedded in regions with density above a certain threshold
summing the dynamical  masses of galaxy groups 
\citep[Table~\ref{tab:gr10} and ][]{2014A&A...566A...1T}.
For poor groups with 
$N_{\mathrm{gal}} \leq 3$ they used 
the median values of masses of poor groups. 
Superclusters contain also  intercluster gas which may form approximately
10\% of their total mass
\citep[see e.g. ][]{2016A&A...592A...6P}. 
Total mass of the overdensity regions in the supercluster,  $M_{\mathrm{tot}}$, 
is obtained by adding this to the dynamical mass. 
This analysis showed that the high density core of SCl~2142 has already started 
to collapse.

For this study we calculated  the mass - radius relation using different
approach. The radial mass distribution is calculated using projected clustercentric
radii. To determine the mass of the supercluster core at a given
clustercentric radius we used group masses embedded into the core at this radius.
For rich groups we use dynamical
masses from Table~\ref{tab:gr10}. We find masses of poor groups 
using stellar mass-halo mass relation. 
According to this, the mass of the haloes can be calculated, employing
the relation between the stellar mass 
$M_{\mbox{*}}$ of the brightest galaxies in a group
to halo mass $M_{\mbox{halo}}$ from \citet{2010ApJ...710..903M},
\begin{equation}
\frac{M_{*}}{M_{\mbox{halo}}}=2\left(\frac{M_{*}}{M_{\mbox{halo}}}\right)_0 
\left[\left(\frac{M_{\mbox{halo}}}{M_1}\right)^
{-\beta}+\left(\frac{M_{\mbox{halo}}}{M_1}\right)^\gamma\right]^{-1},
\label{eq:stmass}
\end{equation}
where $(M_{\mbox{*}}/M_{\mbox{halo}})_0=0.02817$ 
is the normalisation of the stellar to halo mass relation, 
the halo mass $M_{\mbox{halo}}$ is the virial mass of haloes, 
$M_1 = 7.925\times~10^{11}M_\odot$
is a characteristic mass,
and $\beta=1.068$ and $\gamma=0.611$ 
are the slopes of the low- and high-mass ends of the relation, respectively. 
This method was used by  \citet{2016A&A...595A..70E}
to find the mass of the Sloan Great Wall (SGW) superclusters, 
and by \citet{2016A&A...588L...4L} and \citet{2017A&A...603A...5E} to estimate the mass
of the BOSS Great Wall superclusters. 
The total mass was  
obtained summing the masses of groups and by adding 10\% of the mass 
as the mass of gas in the supercluster.

We present the mass-radius relation for the main body of SCl~2142
in Fig.~\ref{fig:massradius}. 
Errors of the mass of a supercluster core at  a given radius 
in Fig.~\ref{fig:massradius} are calculated using
group mass errors, which depend on the group richness
(shaded regions in the figure). 
\citet{2016A&A...595A..70E} estimate that, on average,
supercluster mass errors
due to group dynamical mass errors  are of about 0.3 dex.
For the core region of SCl~A2142 such errors were probably overestimated.
First, 
comparison of different mass estimates for the cluster A2142 
showed that our mass estimate differs 4\% only from those
obtained in \citet{2014A&A...566A..68M}
who compared several mass estimation methods.
We used this as the error estimate of A2142 cluster mass.
Comparison of dynamical mass estimates of groups with those obtained
using stellar masses of groups brightest galaxies showed that in average,
these estimates agree well although individual estimates may
differ up to 50\% \citep{2018A&A...610A..82E}.
Therefore, 
for other rich groups with at least ten members galaxies we used 50\% errors. 
For poor groups, for which we used the stellar
masses of their brightest galaxies to calculate
group masses,   errors are of about 15\%,
obtained using errors of stellar mass estimates of galaxies 
\citep[see Sect. 4.4 in][for details, dark shaded region in the figure]
{2017A&A...603A...5E}.
To compare  different error estimates 
we show in Fig.~\ref{fig:massradius} mass errors
calculated using 50\% errors for both rich and poor groups (except the cluster A2142)
(light shaded region).  
Comparison of masses of SCl~A2142 core regions obtained in this
way (using both stellar masses and group dynamical masses)
with masses calculated using group dynamical masses only
shows that both estimates give very similar results
\citep[][]{2015A&A...580A..69E, 2015A&A...581A.135G}. This
suggests that we have no strong biases in the mass calculations.
The same was concluded in \citet{2016A&A...595A..70E} in which several mass
estimates of the core regions of superclusters from the Sloan Great Wall
were compared.

\begin{figure}[ht]
\centering
\resizebox{0.45\textwidth}{!}{\includegraphics[angle=0]{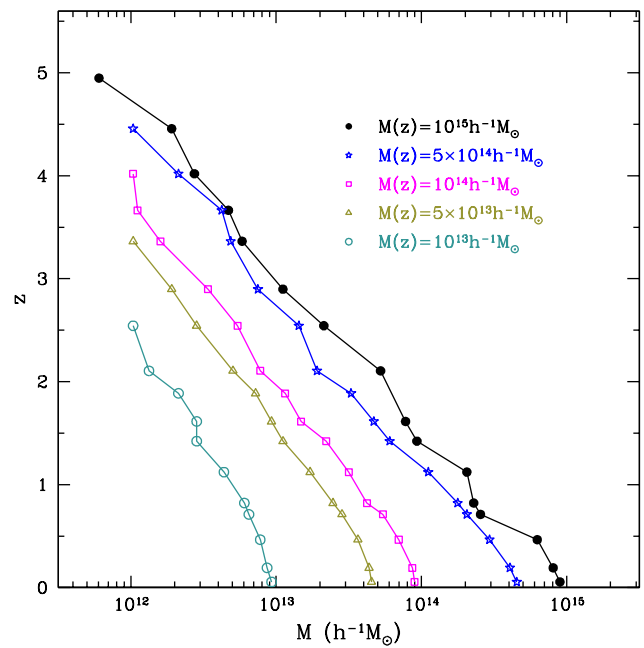}}
\caption{
Evolution of halo mass $M$ with redshift $z$ for five mass samples.
Filled circles show the average
evolutionary track of haloes with the current mass of 
$10^{15}h^{-1}M_\odot \pm 10\%$.
Stars correspond to average mass $5\times~10^{14}h^{-1}M_\odot$,
open squares correspond to mass $10^{14}h^{-1}M_\odot$,
open triangles to mass $5\times~10^{13}h^{-1}M_\odot$, and
open circles to mass  $10^{13}h^{-1}M_\odot$.
}
\label{fig:massevol}
\end{figure}

At clustercentric distances up to
approximately $3$~\Mpc\ the mass of the supercluster core is determined by the
mass of the cluster A2142 and its infalling subclusters, including  group
Gr5 at clustercentric distance of $2.4$~\Mpc. At clustercentric
distances  $3 - 4$~\Mpc\ the masses of groups Gr3 and Gr4 increase
the mass of the supercluster core. According to Fig.~\ref{fig:massradius}, 
the radius of a region which has reached the turnaround and already started
to collapse is of about $8$~\Mpc. 
The total mass of the supercluster core within this
clustercentric distance is $M_{\mathrm{tot}} \approx 2.3\times~10^{15}h^{-1}M_\odot$.
The core of the supercluster with radius of about $9$~\Mpc\ may collapse in the future. 
The total mass in the future collapse region is 
$M_{\mathrm{tot}} \approx 2.4\times~10^{15}h^{-1}M_\odot$.

Using simulations we are able to follow the  evolution of dark matter
haloes \citep[][and references therein]{2015JKAS...48..213K, 2018MNRAS.477.4931H}. 
In Fig.~\ref{fig:massevol} 
we present the mass evolution of simulated haloes found with
FoF method in Horizon Run 4 simulations, as described
in detail in \citet{2015JKAS...48..213K}.  
The mass evolution of haloes with a given mass is plotted 
in Fig.~\ref{fig:massevol} which shows 
the average evolutionary tracks of haloes with a given mass.

This figure shows that the increase of halo mass from high redshifts 
to the present depends on the halo mass. 
For example, the cluster-size haloes with current mass of  $10^{15}h^{-1}M_\odot$ tend to have 
mass of about $10^{13}h^{-1}M_\odot$ at $z = 3$ on average.
The half-mass period of cluster-size haloes 
with the present-day masses of about
$M = 10^{15}h^{-1}M_\odot$ is at redshift $z = 0.5$. 
Approximately, we can use this relation to find how the mass of a collapsing 
core of the supercluster evolves with redshift.
Using Appendix~\ref{sect:sph} and Eq.~\ref{eq:z05} we find the radius
of a collapsing core of the supercluster at this redshift.
The overdensity of the turnaround at $z = 0.5$ is $\Delta\rho_T = 8.0$. 
Using this value, and the half of the mass of the core ($M = 1.15\times~10^{15}h^{-1}M_\odot$), 
we obtain that 
the radius of the collapsing region at $z = 0.5$ was in comoving coordinates
approximately $7.7$~\Mpc.
At redshift $z = 1.0$ the masses of haloes had approximately five times
lower values than their present-day masses.
At this redshift the overdensity at the turnaround is
$\Delta\rho_T = 6.6$ (Appendix~\ref{sect:sph}). 
We find that at redshift $z = 1.0$ the radius of a region which has reached turnaround and
started to collapse was  approximately 
$6$~\Mpc\ (in comoving coordinates). Its mass was $M \approx 5\times~10^{14}h^{-1}M_\odot$.

 
The mass and radius of the collapsing core of SCl~2142
are larger than those in the collapsing cores
in the SGW superclusters.  
In the SGW superclusters the mass of the most massive core which has reached turnaround
is $M \approx 1.8\times~10^{15}h^{-1}M_\odot$ \citep{2016A&A...595A..70E}. 
The radii of collapsing cores in other superclusters 
do not exceed $10$~\Mpc\
\citep{2000AJ....120..523R, 2014MNRAS.441.1601P, 2015MNRAS.453..868O,
2015A&A...575L..14C}. 
\citet{2000AJ....120..523R} and \citet{2014MNRAS.441.1601P} showed that
only very massive collapsing supercluster 
cores with a number of rich galaxy clusters as in
the Shapley supercluster and in the
Corona Borealis supercluster, and perhaps
in the rich proto-supercluster at redshift $z = 2.45$
\citep{2018arXiv180606073C}  may be  larger and more massive.

\subsection{SCl~A2142 and the evolution of clusters and superclusters}
\label{sect:evol}

In this study we 
find that the main cluster of SCl~2142, A2142, and the group Gr8 at the edge
of the supercluster main body are populated by the oldest galaxies among
groups in the supercluster. The median age of galaxies in them is higher than
$8$~Gyr. 
We may suppose that these systems and their galaxies had
already started to form in a very early universe 
together with first galaxies observed at high redshifts 
\citep{2017ApJ...844L..23C, 2018MNRAS.473.2335M, 
2018Natur.553...51M, 2018ApJ...861...43P}.

The radius of a collapsing region around Gr8 with mass of about
$M = 1.6\times~10^{14}h^{-1}M_\odot$ is $R_{\mathrm{T}} \approx 3.3$~\Mpc.
The radius of the collapsing core of A2142 is $R_{\mathrm{T}} \approx 8$~\Mpc. 
As the mass of the cluster A2142 is much higher than that of Gr8,
it is in a much deeper potential well and 
during the evolution, much more galaxies from a larger volume
fall into A2142 than into Gr8 making it richer than Gr8.
Simulations show that present-day clusters 
with mass of about $M \approx 10^{15}h^{-1}M_\odot$ as the cluster A2142
and $M \approx 10^{14}h^{-1}M_\odot$ as the group Gr8
had comoving effective radii at redshift $z = 0.5$ approximately $4$ and $1.5$~\Mpc,
and approximately $10$ and $5$~\Mpc\ at redshift $z = 1$
\citep{2013ApJ...779..127C}. At higher redshifts galaxies may fall into forming clusters
even from larger volumes \citep{2018MNRAS.473.2335M}. 

The collapsing core in the supercluster, 
infalling subclusters and groups around the cluster A2142, and possible group
merging, found in this study and in \citet{2018A&A...610A..82E} show that
the evolution of the supercluster and its  members continues. 
Multi-wavelength studies of the cluster A2142 have revealed
substructure in the cluster, and infalling galaxies and groups 
which affect the galaxy properties in the cluster
\citep{2000ApJ...541..542M, 2014A&A...570A.119E, 2017A&A...605A..25E, 
2017A&A...603A.125V, 2018A&A...610A..82E, 2018ApJ...863..102L}.  
It is possible that during the future evolution all groups in the core
of SCl~A2142 join the main cluster, as predicted by simulations
about the future evolution of superclusters \citep{2009MNRAS.399...97A}.
At present, we evidence one epoch in this evolution.

Galaxy groups in lower density regions have higher fraction of star-forming
galaxies than galaxies in superclusters \citep{2012A&A...545A.104L, 2014A&A...562A..87E}.
Therefore we could expect that the percentage of
star-forming galaxies increases with distance from the supercluster centre
along the supercluster axis. We did not find statistically significant
differences between groups in the supercluster core and in the tail. 
However, the galaxy content of other groups in SCl~2142 depends 
on the location of groups  with respect to the supercluster axis.
Groups on the supercluster axis hosts galaxies that have older galaxy populations than groups
which lie away from the supercluster axis, having median ages 
of about $4 - 6$~Gyr and less than $4$~Gyr, correspondingly.
The supercluster axis present a preferred direction 
for the group and galaxy infall into the main cluster
\citep{2018A&A...610A..82E}. This may affect galaxy content
of groups and cause also an alignment signal
for some infalling groups. In the core of SCl~A2142 the most elongated
group, Gr4, and its brightest galaxy are aligned along the supercluster axis.
Recent discussion and references about the galaxy and cluster
alignments in the cosmic web can be found in \citet{2018A&A...610A..82E},
and in \citet{2018MNRAS.481..414G}.

\citet{2018MNRAS.473L..79B} analysed the galaxy content of
groups infalling into massive clusters at redshifts $0.15 < z < 0.3$.
Their study showed that the star formation quenching is effective
in groups before infalling to clusters. Our study tells us that 
the star formation history in galaxies also depends on the location
of galaxies and groups in the cosmic web. 
Detailed dynamical study with precise peculiar velocity and mass estimates 
are needed to understand the details of the evolution of clusters and superclusters, 
and their dynamics.

\subsection{Star-forming and recently quenched galaxies and AGNs
}
\label{sect:sfshock} 

Star-forming galaxies in groups of SCl~A2142 are located approximately
at the virial radii of groups, or near the edge of groups. In merging groups
these galaxies lie in the zone where galaxies from merging groups 
overlap in the PPS diagram
(see below, and also in Appendix~\ref{sect:gr14}). 
Up to clustercentric distances $4$~\Mpc\ the distribution of clustercentric
distances of star-forming galaxies in rich groups is peaked at higher
distances than that of galaxies with old stellar populations.
We find the enhanced fraction of star forming and recently quenched galaxies 
at clustercentric distances of about  $6 - 8$~\Mpc, especially in poor groups.
This is the borderline of the turnaround region of the supercluster core.
In addition, several AGN host galaxies are recently quenched,
or they lie close to recently quenched galaxies.

In the cold dark matter hierarchical structure formation models, 
galaxy clusters grow by accreting  
matter (galaxies, groups, and gas) from their surroundings. 
Simulations predict the existence of large-scale accretion
shocks between the virial and turnaround radii of clusters of galaxies
\citep{2009ApJ...696.1640M, 2005ApJ...623..632K}.
The densest and hottest shocks occur at the intersections of filaments 
around the locations of clusters of galaxies \citep{2000Natur.405..156L}.
\citet{2014ApJ...781L..40E} found that gas-rich
galaxies possibly shocked by infalling into the cluster or by the
passage of a shock wave can exhibit high star-formation rates. 
Using Fermi-LAT and ROSAT data, signatures 
of virial rings around Coma cluster and
the cumulative $\gamma$-ray emission from accretion shocks
of stacked 112 rich clusters 
were recently detected 
\citep{2017ifs..confE.151K, 2018JCAP...10..010R}.

\citet{2018arXiv180101494K} showed a presence of a  $\gamma$-ray excess
due to the virial shock near the virial radius of the cluster A2142.
\citet{2018A&A...610A..82E} found that star-forming
and recently quenched galaxies in this cluster are located
at the infall region of subclusters, approximately at the same radius 
as detected by \citet{2018arXiv180101494K} as the radius
of a virial shock.  \citet{2018A&A...610A..82E} concluded that
mergers and infall of subclusters affect the properties of galaxies
and lead to rapid changes in galaxy properties. As shown by \citet{2018arXiv180101494K},
this is accompanied by a virial ring at high energies. 

Several AGNs lie at the  clustercentric distances of about $4$~\Mpc.
Among them there is one FRII galaxy.  
At these distances we did not found any recently quenched galaxy.
\citet{2015MNRAS.450..646S} and \citet{2015MNRAS.450..630S}
found that shock waves in merging clusters  can trigger the
star formation and AGN activity 
in galaxies for a short time, of order of 100~Myr.
\citet{2008ApJ...676..147B} showed that average FRII 
lifetime in groups is even shorter, $15(\pm 5)\times~10^{6}$~years. 
The absence of recently quenched galaxies at
$D_c \approx 4$~\Mpc\ may be due to the short timescale at which shock waves
trigger the star formation in galaxies 
\citep{2015MNRAS.450..646S, 2015MNRAS.450..630S}.  
Radio lobes of the FRII galaxy in Gr3 are
aligned along the supercluster axis
\citep[see Fig. A.1 in ][]{2017A&A...601A..81C}.
However, this may be a
coincidence. To understand this, studies of the
co-evolution of the structures of different scales from active nuclei in galaxies
to groups and superclusters of galaxies are needed.

The distributions of clustercentric distances
of galaxies with old and young populations in groups are shifted,
galaxies with old stellar populations lie at lower clustercentric
distances than galaxies with young stellar populations. This 
may be an indication that stellar populations in galaxies closer to the main cluster
were quenched earlier than those in galaxies farther away.

In groups Gr8, Gr12, and Gr13 (Figs.~\ref{fig:gr6885pps} -  \ref{fig:gr10224pps})
star-forming  galaxies and AGNs are located approximately
at the virial radius of a group, near the edge of a group, or 
they lie in the zone where galaxies from possibly merging groups 
overlap. 
\citet{2018A&A...610A..82E} found that in the cluster A2142 
star-forming galaxies lie in the infall zone of subclusters. 
This has also been found in other studies. For example, 
\citet{2017A&A...607A.131D} have
detected a high percentage of star-forming galaxies in the cluster A520
in infalling groups at high clustercentric distances.
\citet{2009A&A...500..947B} and \citet{2012MNRAS.427.1252M} showed that in 
dynamically unrelaxed clusters the average star formation
history of cluster member galaxies  depends both  on clustercentric distance 
of galaxies, and on cluster substructure.
\citet{2017ApJ...841...18B} found that colours of galaxies from the SDSS 
change at the halo boundaries.
This may be evidence of merger-induced star formation in
galaxies. \citet{2016arXiv160707881A} associate the changes in star formation properties 
of galaxies with cosmic web detachment as galaxies fall into clusters.
 
Possible merger-induced star formation in galaxies along filaments between clusters
have been reported by \citet{2008MNRAS.390..289J} who found 
an excess of  blue star-forming galaxies in the filament which connects clusters 
in the core region of the Horologium-Reticulum supercluster. 
An excess of 
star-forming  galaxies have also been found in filaments surrounding
other galaxy clusters \citep{2007MNRAS.375.1409P, 2007A&A...470..425B,
2008MNRAS.388.1152P, 2008ApJ...672L...9F, 2010AJ....140.1891E}.

We find an excess of star-forming galaxies,
recently quenched galaxies,  
and AGNs at clustercentric distances of about  $6$~\Mpc.
This is more clearly seen in the distribution of galaxies 
from poor groups. At this distance 
star-forming and recently quenched galaxies are spread over the whole 
main body of the supercluster. They are not located preferentially
along the supercluster axis or in filaments in the supercluster
main body. We could expect that star-forming galaxies lie
in the outer parts of the supercluster main body, so that
the fraction of star-forming galaxies increases with clustercentric distance.
The increase of the fraction of star-forming galaxies around clusters
have been found in several other studies  \citep{2013MNRAS.430.3017B, 2015ApJ...806..101H}.
Opposite to this, in outskirts of SCl~A2142 main body the fraction of
star-forming galaxies decreases. 
The origin of the excess at $D_c \approx 6$~\Mpc\ is not yet clear.

\section{Summary}
\label{sect:sum} 

We analysed the structure, dynamical state, and galaxy content
of the supercluster SCl~2142.  
Our main results are as follows.

\begin{itemize}
\item[1)]
The total mass of the collapsing core of the supercluster 
is $M_{\mathrm{tot}} \approx 2.3\times~10^{15}h^{-1}M_\odot$, and
its radius is approximately $8$~\Mpc. 
Groups in the supercluster core up to 
clustercentric distances of about $8$~\Mpc\ are infalling into the main cluster.
\item[2)]
Stellar populations in galaxies in groups on the supercluster axis are older than 
in off-axis groups,
with median stellar ages of $4 - 6$ (on-axis groups) and $< 4$~Gyr (off-axis groups). 
Populations in poor groups have median ages of order of $3$~Gyr.
Stellar populations are the oldest (with median stellar age $t > 8$~Gyr) 
in galaxies in the central
cluster, A2142, and group Gr8  at the clustercentric distance of about $10$~\Mpc.
\item[3)]
Group Gr5 with its small clustercentric distance and  
galaxy populations similar to populations in the cluster A2142 and its infalling
subcluster
may be a part of a small filament infalling into the cluster A2142.
\item[4)]
The small magnitude gap between the brightest galaxies in groups, 
substructure, infall, and merging  of groups suggest 
that galaxy groups in SCl~A2142 are dynamically young.
\item[5)]
There is  an excess of star-forming and
recently quenched galaxies and AGNs in poor groups 
at the clustercentric distances 
$D_c \approx 6$~\Mpc. 
\end{itemize}

The progenitors of the cluster A2142 
might be similar to rare, very rich protoclusters 
which are now observed at high redshifts.
The cluster is embedded in
a supercluster with unusual morphology among superclusters, 
having almost spherical, collapsing
main body and straight tail \citep{2015A&A...580A..69E}.
The collapsing core of the supercluster,
 dynamically young galaxy groups in it, 
infall of galaxies and groups, and their possible 
merging which affect galaxy properties, as found in this and earlier studies
(see references in Sect.~\ref{sect:evol}), 
show how the whole supercluster
is evolving. To check whether the abundance and evolution of very rich clusters
and superclusters
is compatible with the current cosmological model we need to analyse
halo properties in very large-volume simulation like the Horizon Run 4 simulations
\citep{2015JKAS...48..213K}. 
It is necessary to analyse a large sample of galaxy superclusters and 
galaxies and groups in them to better understand the coevolution
of different structures in the cosmic web.

\begin{acknowledgements}

We thank the referee for suggestions and comments that helped to improve the paper. 
We are pleased to thank the SDSS Team for the publicly available data
releases.  Funding for the Sloan Digital Sky Survey (SDSS) and SDSS-II has been
provided by the Alfred P. Sloan Foundation, the Participating Institutions,
the National Science Foundation, the U.S.  Department of Energy, the
National Aeronautics and Space Administration, the Japanese Monbukagakusho,
and the Max Planck Society, and the Higher Education Funding Council for
England.  The SDSS website is \texttt{http://www.sdss.org/}.
The SDSS is managed by the Astrophysical Research Consortium (ARC) for the
Participating Institutions.  The Participating Institutions are the American
Museum of Natural History, Astrophysical Institute Potsdam, University of
Basel, University of Cambridge, Case Western Reserve University, The
University of Chicago, Drexel University, Fermilab, the Institute for
Advanced Study, the Japan Participation Group, The Johns Hopkins University,
the Joint Institute for Nuclear Astrophysics, the Kavli Institute for
Particle Astrophysics and Cosmology, the Korean Scientist Group, the Chinese
Academy of Sciences (LAMOST), Los Alamos National Laboratory, the
Max-Planck-Institute for Astronomy (MPIA), the Max-Planck-Institute for
Astrophysics (MPA), New Mexico State University, Ohio State University,
University of Pittsburgh, University of Portsmouth, Princeton University,
the United States Naval Observatory, and the University of Washington.

The present study was supported by the ETAG projects 
IUT26-2 and IUT40-2, by the European Structural Funds
grant for the Centre of Excellence "Dark Matter in (Astro)particle Physics and
Cosmology" TK133, and by grant MOBTP86. 
HL is funded by PUT1627 grant from Estonian Research Council.
This work was partially supported by the Supercomputing Center/Korea
Institute of Science and Technology Information with supercomputing
resources including technical support (KSC-2013-G2-003). JK and CP thank
Korea Institute for Advanced Study for providing computing resources (KIAS
Center for Advanced Computation) for the analysis of the HR4 data.
This work has also been supported by
ICRAnet through a professorship for Jaan Einasto.

\end{acknowledgements}

\bibliographystyle{aa}
\bibliography{scl2142core.bib}

\begin{appendix}

\section{Individual galaxy groups}
\label{sect:gr14} 

\begin{figure}[ht]
\centering
\resizebox{0.40\textwidth}{!}{\includegraphics[angle=0]{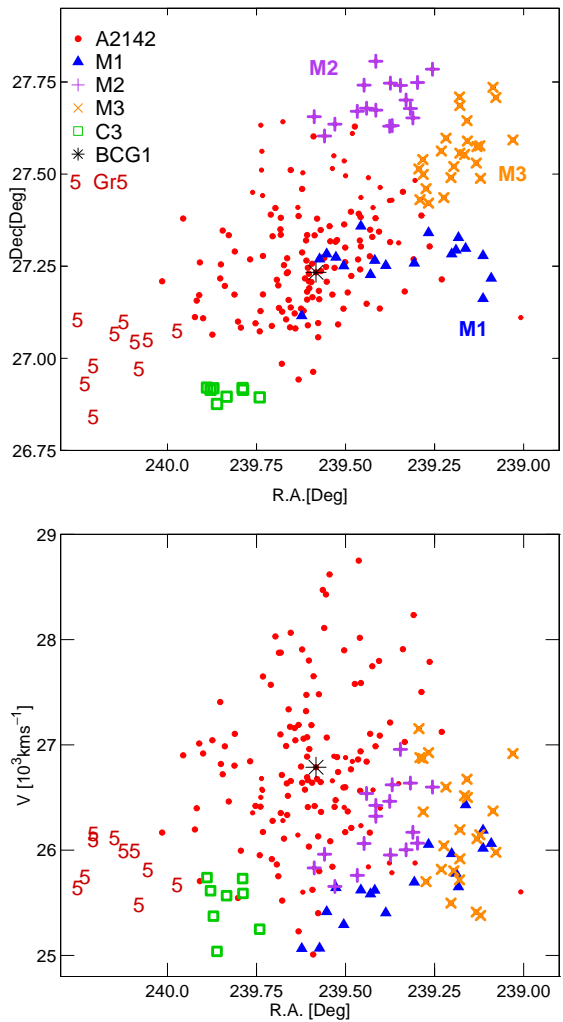}}
\caption{
Distribution of galaxies in three components of the cluster A2142 and in the group Gr5
in the sky plane (upper panel)
and in the R.A. - velocity plane (lower panel).
Red symbols correspond to the galaxies in the main component of A2142, C1.
Blue, violet, and orange symbols show galaxies from infalling
subclusters M1, M2, and M3 (see text), and 
green symbols denote the location of galaxies from the third
component, C3. Black star indicates the location of the brightest cluster 
galaxy (BCG1).
}
\label{fig:a2142gr5}
\end{figure}

Below we describe the properties of individual groups
with at least ten member galaxies, their
substructure, dynamical state, orientation,  the location of star forming
galaxies and AGNs in a group, and possible group merging.
For each group we present an analysis relevant to
this particular group. For example, we search for possible substructure 
in groups with more than 12 galaxies only, for groups with smaller
number of galaxies the results of {\it mclust} are less reliable,
as shown by \citet{2013MNRAS.434..784R}. If we identify
AGNs in a group we note their location, but AGNs have not been found
 in every group.
For two groups we present the PPS diagrams with galaxies from neighbouring poor 
groups and discuss the possibility that groups may be merging.

{\bf Group Gr5}, 
a small  group of ten galaxies, is the closest group to the main cluster
of the supercluster, A2142.  
\citet{2018A&A...610A..82E} found that the cluster A2142 has three infalling
subclusters, 
named M1, M2, and M3, and a poor infalling group 
called as C3. 
Subclusters M1-M3 may correspond to substructures identified
in the cluster A2142 by \citet{2011ApJ...741..122O}, as discussed
also in \citet{2018A&A...610A..82E}, and recently in
\citet{2018ApJ...863..102L}. We refer to these papers
for details about substructures in the cluster A2142.  
In Fig.~\ref{fig:a2142gr5} we show the distribution of Gr5 galaxies
together with galaxies from A2142 cluster and its infalling subclusters
and group (M1 - M3, and C3)
in the sky plane and in the sky-velocity plane.
This figure shows that the group Gr5 lies as close to the main cluster as its 
infalling subclusters. Galaxy populations in Gr5
are very similar to those in an infalling subcluster M1 embedding galaxies
with old, passive stellar populations \citep{2018A&A...610A..82E},
with median values of the $D_n(4000)$ index $1.8$ (Gr5), and 1.9 (M1).
The group G5 and subcluster M1 may be parts of a small infalling filament.

{\bf Group Gr3} is the 
richest galaxy group in the supercluster  main body
with 54 member galaxies near the cluster A2142
at median clustercentric distance of $3.3$~\Mpc. The analysis of the
structure of Gr3 with {\it Mclust}, using as input sky coordinates
and velocities of it member galaxies,  
showed that the group consists of  three components.
Such model has the highest value of BIC among all models
considered by {\it mclust}.
In this case the mean uncertainty of galaxies to belong to components
was less than $10^{-4}$. One-component model is not favoured
by {\it mclust}. According to BIC values,
the best such model (with the highest BIC value of the one-component models)
is in 17th place among all models.

\begin{figure}[ht]
\centering
\resizebox{0.45\textwidth}{!}{\includegraphics[angle=0]{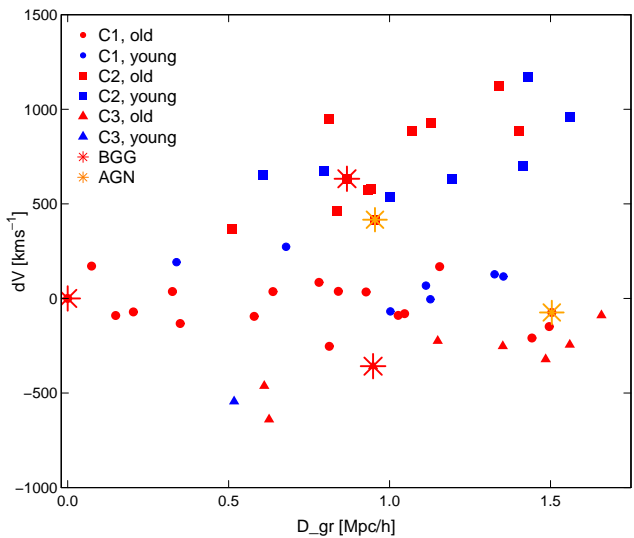}}
\caption{
Velocity of galaxies with respect to the group mean velocity vs. 
the projected distance from the group centre  
$D_{\mathrm{gr}} $ at the brightest galaxy
of the first component in Gr3 (BGG1)
for the components of the group Gr3.
Red symbols show galaxies with old stellar populations
($D_n(4000) \geq 1.55$), and blue symbols denote galaxies with young 
stellar populations ($D_n(4000) < 1.55$).
Filled circles denote galaxies from the component C1, squares
from component C2, and triangles from component C3. 
Red stars show the location of the BGG of each component, and orange stars
correspond to AGNs. 
}
\label{fig:gr4952comppps}
\end{figure}

We present the PPS diagram   
of Gr3 galaxies in three components in Fig.~\ref{fig:gr4952comppps}.
In both figures galaxies with old stellar populations (divided using
$D_n(4000)$ index) are plotted with red symbols and galaxies 
with young stellar populations with blue symbols.

The first component of Gr3, C1, has
26 galaxies.  Figure~\ref{fig:gr4952comppps} shows that
within  the projected distance from the group centre at the brightest
group galaxy (BGG1)
up  to  $D_{\mathrm{gr}} \approx 0.5$~\Mpc\ 
all galaxies (with one exception) in the group
have old stellar populations with ages over $8$~Gyr. 
They are almost all at the same redshift, with
the velocity differences less than $250$~km $s^{-1}$.
The brightest galaxy in this group is FRII radio galaxy.
Most galaxies with young stellar populations lie at 
distances $D_{\mathrm{gr}} > 1.0~$~\Mpc\ from the brightest galaxy. 
The radio lobes of this galaxy are aligned with
the supercluster axis. 

The second component of Gr3, C2 with 19 galaxies lies along the supercluster axis
with position angle of $73 \pm 5^{\circ}$.
Figure~\ref{fig:gr4952comppps} shows that galaxies in this
component have positive velocities in respect of the main component
of the group. They may not lie in the infall zone of C1. 

Galaxies from the third component C3 with nine member galaxies
have negative velocities
in respect of the first component, and they lie in the  infall zone of C1.
The component C3 embeds one luminous galaxy with very old stellar
populations (with ages over $8$~Gyr) only, resembling
typical poor galaxy group with bright main galaxy and fainter satellites.

The difference in galaxy velocities supports the possibility that
C2 is a small separate group on the supercluster axis. 
It is combined together
with the first component because of high density of galaxies
and groups in the supercluster core where the criteria to define 
galaxy groups used in \citet{2014A&A...566A...1T} may join different groups.
Small magnitude gap between the luminosities of the brightest galaxies
in the group shows that the group is still forming.

\citet{2008ApJ...676..147B} shows that redder
galaxies have an increased probability of embedding radio
sources, and the radio fraction increases with the luminosity of host
galaxy. This may be why the brightest galaxy in a group is FRII radio galaxy.

{\bf Group Gr4}
is a group of 11 galaxies with very elongated shape on
the supercluster axis. Gr4 is aligned
along the supercluster axis with position angle of $54 \pm 7^{\circ}$.
The brightest galaxy of this group lies in the centre of the group,
this is one of AGNs in the supercluster, and
faint radio source according to 
VLA FIRST survey data at 1.4 GHz. Its low flux level (19.5 mJy)
suggests  that the  radio emission most likely comes from star formation processes.
The brightest galaxy in the group 
is also aligned  along the supercluster axis.
In this group all galaxies closer to the main cluster than the brightest galaxy 
have old
stellar populations. All galaxies with young stellar populations
lie farther away from the central cluster.
We may suppose that the shape and 
orientation of the group and its BGG, as well as the stellar populations
of galaxies in the group are affected by the infall into the main cluster. 

{\bf Groups Gr6 and Gr7}
lie close together along the supercluster axis at clustercentric distances
of $7$~\Mpc\ and $6$~\Mpc, correspondingly, out of the 
collapsing core of the supercluster.  The Gr7 
embeds one of the AGNs in the supercluster, which is also a faint
radio source. It also has one recently quenched galaxy, located near virial radius
of the group. 
The Gr7 group hosts three galaxies with very old stellar populations having
ages $t > 9.5$~Gyr, while Gr6 has seven galaxies with such old 
stellar populations, and no AGNs and recently quenched galaxies. 
The supercluster axis is probably the preferred
direction of galaxy and group infall into the main cluster  
\citep{2017A&A...603A.125V, 2018A&A...610A..82E}. On this axis 
Gr7 is closer to the main cluster than Gr6,
and AGN phenomenon and stellar populations of galaxies in this group 
may be affected by this.

{\bf Groups Gr1 and Gr9} are 
groups  of 27 and 21 galaxies, respectively. They are located 
away from the supercluster axis, at the 
clustercentric distance of $4.4$~\Mpc\ and $5.4$~\Mpc.
In the PPS diagram (Fig.~\ref{fig:a2142cdv8}) galaxies from Gr1 have
the highest positive values of their velocities in respect of the cluster
centre. 
In Gr9 the third brightest
galaxy is an AGN. Both Gr1 and Gr9 host one recently quenched galaxy.  
Galaxies in these groups have younger stellar populations that any group
in the supercluster main body at the supercluster axis,
having median ages $t < 4$~Gyr.

\begin{figure}[ht]
\centering
\resizebox{0.48\textwidth}{!}{\includegraphics[angle=0]{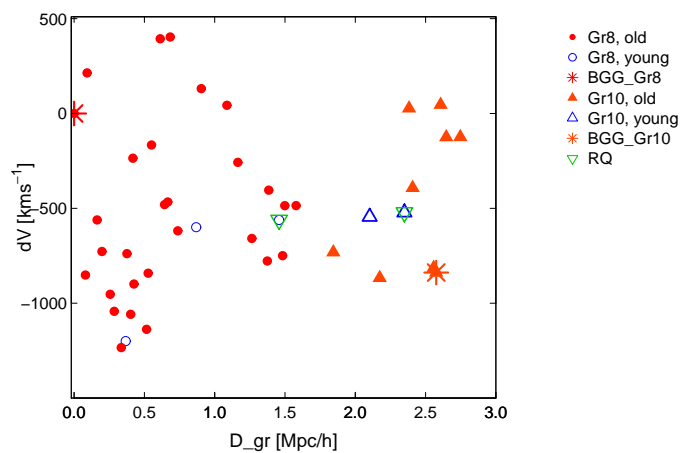}}
\caption{
Velocity of galaxies with respect to the cluster mean velocity vs. 
the projected distance from the group Gr8 centre  
$D_{\mathrm{gr}}$  for the galaxies from Gr8 with Gr10.
Filled red symbols show galaxies with old stellar populations
($D_n(4000) \geq 1.55$), and empty blue symbols denote galaxies with young 
stellar populations ($D_n(4000) < 1.55$).
Green triangles indicate recently quenched galaxies (see text for definition).
Red stars show the location of the BGGs.
}
\label{fig:gr6885pps}
\end{figure}

{\bf Group Gr8}
is  a rich group with 32 member galaxies on the supercluster
axis at the distance of almost $10$~\Mpc\ from the 
supercluster centre. It is aligned along the supercluster axis, with
position angle of $55 \pm 13^{\circ}$.
In Fig.~\ref{fig:gr6885pps} we plot the PPS diagram for this group,
together with neighbouring group of 11 galaxies, Gr10.
Figure~\ref{fig:gr6885pps} suggests that 80\%
of galaxies in Gr8
have negative velocity offsets in respect of the main galaxy
of Gr8. 
These galaxies may be recently accreted into the cluster, 
or still infalling 
\citep[we refer to Sect.~\ref{sect:ppsm} about the possible
interpretation of the PPS diagram, and also to
Fig. 13 in ][]{2015ApJ...806..101H}.
It is also possible that the brightest galaxy in this group has positive
velocity offset
in respect to the group centre.

Galaxies from Gr10 have negative velocity offsets with respect to the main galaxy
of Gr8, so that they lie in the infall zone of Gr8,
at distances approximately $2 - 3$~\Mpc\ from the Gr8 brightest
galaxy. 
This can be interpreted so that group Gr10 is falling into the Gr8
along the supercluster axis (see also the sky distribution of galaxies
in these groups in Figs.~\ref{fig:radec} and \ref{fig:radecfro}). 
The radius of a collapsing region around Gr8 is $R_{\mathrm{T}} \approx 3.3$~\Mpc\
(Sect.~\ref{sect:evol}), thus galaxies from Gr10 are embedded 
in this region. 
It is also possible that Gr8 and Gr10 already form a single group,
elongated along the supercluster axis,
and this group is split into separate systems by our group finding algorithm.
Figure~\ref{fig:gr6885pps} shows that 
the central part of the group up to its virial radius, $\approx 0.5$~\Mpc\ is populated
with galaxies with old stellar populations. 
Six out of nine star-forming galaxies with young
stellar populations in groups Gr8 and Gr10, and
recently quenched galaxies lie in the 
the projected distance
interval from the group centre approximately $1 - 2.5$~\Mpc, where groups may merge.
 It is possible
that star formation in these galaxies is triggered by group merger.

\begin{figure}[ht]
\centering
\resizebox{0.45\textwidth}{!}{\includegraphics[angle=0]{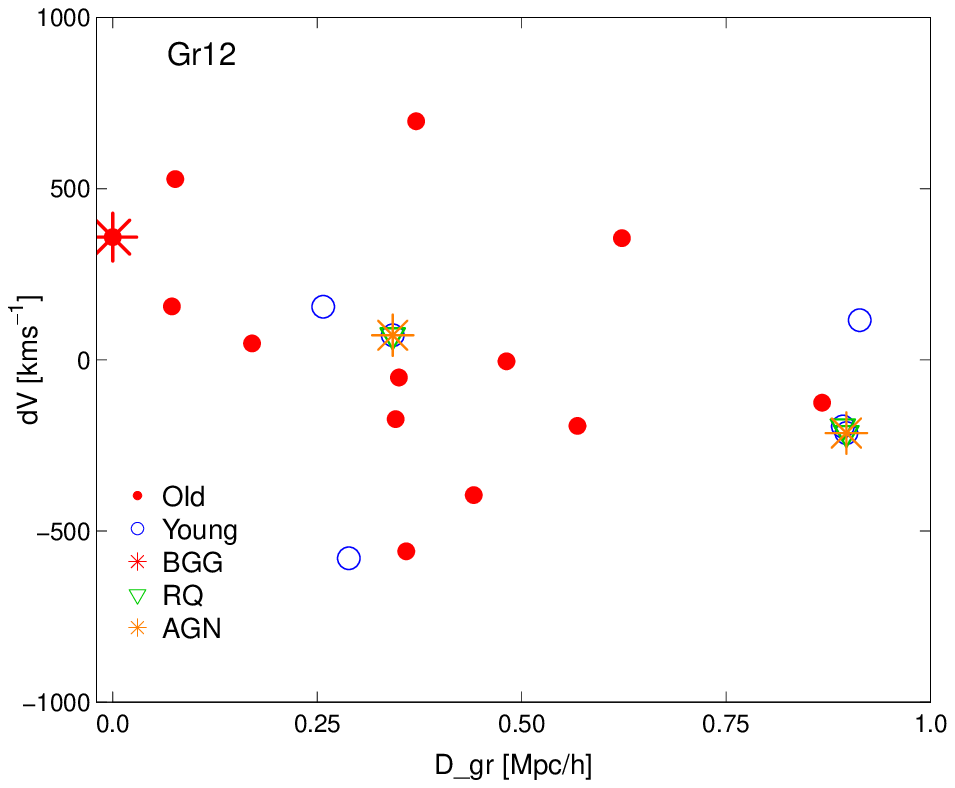}}
\caption{
Velocity of galaxies with respect to the cluster mean velocity vs. 
 the projected distance from the group centre  
$D_{\mathrm{gr}}$   for group Gr12 galaxies.
Filled red symbols show galaxies with old stellar populations
($D_n(4000) \geq 1.55$), and empty blue symbols denote galaxies with young 
stellar populations ($D_n(4000) < 1.55$).
Green triangles indicate recently quenched galaxies, and orange stars denote
AGNs. Red star shows the location of the BGG.
}
\label{fig:gr14283pps}
\end{figure}

{\bf The group Gr12} with 19 member galaxies (possibly up to 25,
if we include galaxies omitted because of fibre collisions) in the supercluster tail
hosts two AGNs and even three recently quenched galaxies.
This is the largest number of recently quenched galaxies among rich groups 
in the supercluster. This group  hosts the lowest
percentage of old, passive galaxies among groups on the supercluster axis. 
In Fig.~\ref{fig:gr14283pps} we plot the PPS diagram for galaxies in this group.
It shows that 
galaxies with old stellar populations, AGNs,
and recently quenched galaxies lie at the projected distance from the group centre  
$D_{\mathrm{gr}}$  larger than  $0.25$~\Mpc, 
which is close to the virial radius of the group,
and at the borders of the group.

\begin{figure}[ht]
\centering
\resizebox{0.45\textwidth}{!}{\includegraphics[angle=0]{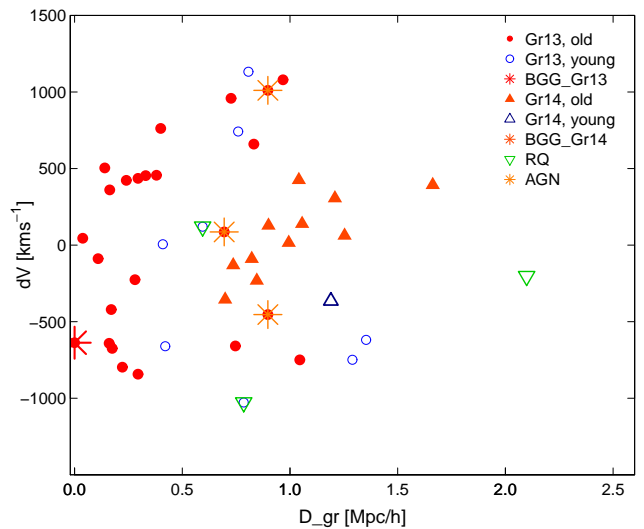}}
\caption{
Velocity of galaxies with respect to the cluster mean velocity vs. 
the projected distance from the group Gr13 centre  
$D_{\mathrm{gr}}$  for the galaxies from Gr13 with Gr14.
Notations as in Fig.~\ref{fig:gr14283pps}.
}
\label{fig:gr10224pps}
\end{figure}

{\bf Groups Gr13 and Gr14} lie at the edge of the supercluster. 
In Fig.~\ref{fig:gr10224pps} we present the PPS diagram 
of groups Gr13 and Gr14.
This figure shows that galaxies from both groups lie at overlapping regions
of the PPS diagram suggesting that these groups are  merging.
This agrees with the prediction of the spherical 
collapse model which tells that the mass of these groups is
enough to become a collapsed system \citep{2015A&A...580A..69E}. 
Gr13 and Gr14  may already form a single group,
which is split into separate systems by our group finding algorithm.
At the projected distances from the group centre  
$D_{\mathrm{gr}}$   up to $0.5$~\Mpc\ (approximately
the virial radius of the group) all galaxies in Gr13 have 
old stellar populations. Almost all galaxies that have young
stellar populations lie in the overlapping region of the two galaxy groups,
at the projected distances from the group centre approximately $0.5 - 1.0$~\Mpc. 
Here lie also recently quenched galaxies and AGNs.
In Gr14 the median age of stellar populations of galaxies is the lowest
among groups in SCl~2142. 
Star formation, ages of stellar populations, 
and recent quenching in these galaxies may be triggered by group merger,
which could also trigger the activity of AGNs.

\section{Spherical collapse model}
\label{sect:sph}

The spherical collapse model describes the evolution of a spherical 
perturbation in an expanding universe. 
This model was studied, for example,  by
\citet{1980lssu.book.....P},
\citet{1984ApJ...284..439P}, and \citet{1991MNRAS.251..128L}.  
In the standard models with cosmological constant, 
the dark energy started accelerating the expansion at the redshift 
$z \approx 0.7$ 
and the formation of structure slowed down. 
At the present epoch, the largest bound structures are just forming.  
In the future evolution of the universe, 
these bound systems separate from each other at an accelerating rate, 
forming isolated `island universes'
\citep{2002PhRvD..65l3518C, 2006MNRAS.366..803D}.  

Under the assumption of sphericity, 
the dynamics of a collapsing shell is determined 
by the mass in its interior.
The evolution of a collapsing shell
have several epochs which can be characterised by density contrasts
\citep{2015A&A...575L..14C, 2015A&A...581A.135G}.
These density contrasts can be used to derive the relations 
between radius of a perturbation and the interior mass for each essential epoch.

For a spherical volume 
the density ratio to the mean density (overdensity)
$\Delta\rho = \rho/\rho_{\mathrm{m}}$ can be calculated as in
 \citet{2016A&A...595A..70E} 
\begin{equation}
\Delta\rho=6.88\,\Omega_\mathrm{m}^{-1}\left(\frac{M}{10^{15}h^{-1}M_\odot}\right)
        \left(\frac{R}{5h^{-1}\mathrm{Mpc}}\right)^{-3}.
\label{eq:sph}
\end{equation}
From Eq.~(\ref{eq:sph}) we can find the mass of a structure as
\begin{equation}
M(R)=1.45\cdot10^{14}\,\Omega_\mathrm{m}\Delta\rho\left(R/5h^{-1}\mathrm{Mpc}\right)^3h^{-1}M_\odot.
\label{eq:mass1}
\end{equation}


{\it Turnaround.} One essential moment in the evolution of a spherical perturbation 
is called turnaround, the moment when the sphere stops expanding 
together with the universe and the collapse begins. 
At the turnaround, the perturbation decouples 
entirely from the Hubble flow of the homogeneous background. 
The spherically averaged radial velocity around a system 
in the shell of radius $R$ can be written as $u = HR - v_{\mathrm{pec}}$, 
where $v_{\mathrm{H}} = HR$ is the Hubble expansion velocity and 
$v_{\mathrm{pec}}$ is 
the averaged radial peculiar velocity towards the centre of the system. 
At the turnaround, the peculiar velocity $v_{\mathrm{pec}} = HR$ and $u = 0$. 

The peculiar velocity $v_{\mathrm{pec}}$ 
is directly related to the overdensity $\Delta\rho$.
In the model with a critical mass density, the overdensity 
at the turnaround is $\Delta\rho_{\mathrm{T}} =  5.55$ and it does not change with 
time. In the standard models with 
cosmological constant, the characteristic density contrasts increase during the
evolution. For $\Omega_{\mathrm{m}} = 0.27$ and $\Omega_{\mathrm{\Lambda}} = 0.73$  
the overdensity at the turnaround at redshift $z = 0$ 
is  $\Delta\rho_{\mathrm{T}} =  13.1$ and  
the mass of a structure  at the turnaround  is \citep{2015A&A...581A.135G}
\begin{equation}
M_\mathrm{T}(R)=5.1\cdot10^{14}\left(R/5h^{-1}\mathrm{Mpc}\right)^3h^{-1}M_\odot.
\label{eq:mrt}
\end{equation}

At the redshift $z = 0.5$,
the overdensity at the turnaround  in the spherical collapse model is
$\Delta\rho_T = 8.0$ and the mass of a structure at the turnaround  is
\begin{equation}
$$
M_\mathrm{T}(R)=3.1\cdot10^{14}\left(R/5h^{-1}\mathrm{Mpc}\right)^3h^{-1}M_\odot.
$$
\label{eq:z05}
\end{equation}

At the redshift $z = 1.0$, the overdensity at the turnaround is
$\Delta\rho_T = 6.6$. 


{\it Future collapse.} The superclusters that have not reached the turnaround at present may 
eventually turnaround and collapse in the future \citep{2006MNRAS.366..803D}. 
\citet{2015A&A...575L..14C} showed that for $\Omega_{\mathrm{m}} = 0.27$
the overdensity  
for the future collapse 
 $\Delta\rho_{\mathrm{FC}} =  8.73,$ which gives
the minimum mass of the structure that will 
turn around and collapse in the future as
\begin{equation}
M_\mathrm{FC}(R)=3.4\cdot10^{14}\left(R/5h^{-1}\mathrm{Mpc}\right)^3h^{-1}M_\odot.
\label{eq:mrfvs}
\end{equation}

\begin{figure}[ht]
\centering
\resizebox{0.45\textwidth}{!}{\includegraphics[angle=0]{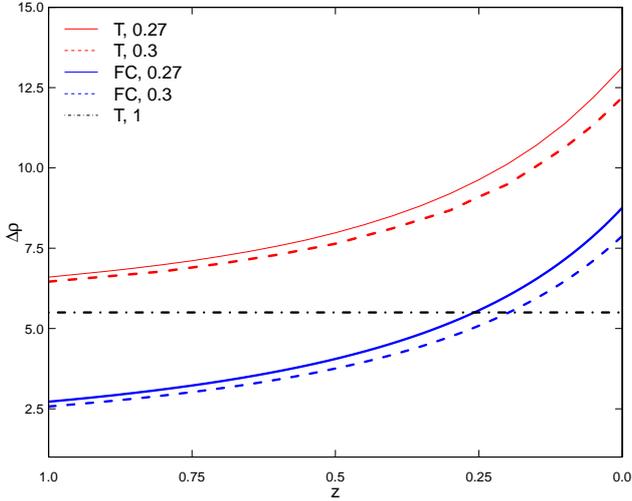}}
\caption{
Evolution of
characteristic density contrasts for the turnaround 
and future collapse with redshift $z$. The solid lines
correspond to $\Omega_m=0.27$ and dashed lines to $\Omega_m=0.3$.
Red lines show turnaround, and blue lines show future collapse.
The dot-dashed line corresponds to the universe with a critical mass density.}
\label{fig:turn}
\end{figure}

Figure~\ref{fig:turn} shows the overdensities for the turnaround 
and future collapse from redshift $z = 1.0$ to the present,
$z = 0.0$, in the standard model with
$\Omega_{\mathrm{m}} = 0.27$ and $\Omega_{\mathrm{\Lambda}} = 0.73$,
in the model with $\Omega_{\mathrm{m}} = 0.3$ and $\Omega_{\mathrm{\Lambda}} = 0.7$,
and in the model with a critical mass density. In this model the overdensity 
at the turnaround is $\Delta\rho_{\mathrm{T}} =  5.55$ and it does not change with 
time.

\end{appendix}

\end{document}